\documentclass[preprint]{aastex}
\usepackage{amsfonts}
\usepackage{amssymb}
\usepackage{subfigure}
\usepackage{graphicx,color}

\newcommand{\psec}{\ensuremath{\, {\rm s}^{-1}}}

\newcommand{\cm}{\ensuremath{\,{\rm cm}}}

\newcommand{\km}{\ensuremath{\,{\rm km}}}

\newcommand{\Mpc}{\ensuremath{\,{\rm Mpc}}}
\newcommand{\K}{\ensuremath{\, {\rm K}}}

\begin{document}

\title{{\LARGE An analytical model of the large neutral regions during the late stage of reionization}}
\author{{\large Yidong Xu\altaffilmark{1}, Bin Yue\altaffilmark{1}, 
Meng Su\altaffilmark{2,3,4}, Zuhui Fan\altaffilmark{5}, Xuelei Chen\altaffilmark{1,6}
\\}}

\altaffiltext{1}{National Astronomical Observatories, Chinese Academy of Sciences, Beijing 100012, China}
\altaffiltext{2}{Department of Physics, and Kavli Institute for Astrophysics and 
Space Research, Massachusetts Institute of Technology, Cambridge, MA 02139, USA}
\altaffiltext{3}{Institute for Theory and Computation, Harvard-Smithsonian Center for 
Astrophysics, 60 Garden Street, MS-51, Cambridge, MA 02138, USA}
\altaffiltext{4}{Eistein Fellow}
\altaffiltext{5}{Department of Astronomy, Peking University, Beijing 100871, China}
\altaffiltext{6}{Center for High Energy Physics, Peking University, Beijing 100871, China}

\begin{abstract}
In this paper we investigate the nature and distribution of large 
neutral regions during the late epoch of reionization.  In the 
``bubble model'' of reionization, the mass distribution of large 
ionized regions (``bubbles'') during the early stage of reionization
is obtained by using the excursion set model, where  the ionization
of a region corresponds to  the first up-crossing of a barrier by 
random trajectories. We generalize this idea, and develop a method 
to predict the distribution of large scale  neutral regions during 
the late stage of  reionization, taking into account the ionizing 
background after the  percolation of HII regions. The large scale 
neutral regions  which we call ``neutral islands'' are not individual 
galaxies or minihalos, but  larger regions where fewer galaxies formed 
and hence ionized later, and they are identified in the excursion set 
model  with the first down-crossings of the island barrier.  Assuming 
that the consumption rate of  ionizing background photons is proportional 
to the surface area of the  neutral islands, we  obtained the size 
distribution of the neutral islands. We also take the ``bubbles-in-island''  
effect into account by considering the conditional probability of  
up-crossing a bubble barrier after down-crossing the island barrier. 
We find  that this effect is very important. An additional barrier is 
set to avoid islands being percolated through. We find that there is 
a characteristic scale for the  neutral islands, while the small 
islands are rapidly swallowed up by the ionizing background, this 
characteristic scale does not change much as the reionization  proceeds.
\end{abstract}

\keywords{Cosmology: theory --- dark ages, reionization, first stars --- 
intergalactic medium --- large-scale structure of Universe --- Methods: analytical}

\section{Introduction}\label{Intro}

The cosmic reionization is one of the most important but poorly understood 
epochs in the history of the Universe. As the first stars form in the 
earliest non-linear structures, they illuminate the ambient intergalactic 
medium (IGM), create HII regions around them, and start the reionization 
process of hydrogen. As the sources become brighter and more numerous, 
HII regions grow in number and size, then merge with each other 
and eventually percolate throughout the IGM.
Various observations have put constraints on the reionization process.
Based on an instantaneous reionization model, the temperature and 
polarization data of the 
cosmic microwave background (CMB) constrain the redshift of reionization 
to be $z_{\rm reion}= 11.1\pm 1.1$ ($1\sigma$, \citealt{2013arXiv1303.5076P}), 
while 
the absence of
Gunn-Peterson troughs \citep{1965ApJ...142.1633G} in high redshift quasar (QSO) 
absorption spectra suggest that the reionization of hydrogen was very 
nearly complete by $z \approx 6$ (e.g. \citealt{2006ARA&A..44..415F}). 
Several deep extra-galactic surveys have found more than 200 galaxies 
at $z \sim 7-8$, but these are still the tip of iceberg, i.e. the most 
luminous of the galaxy population at those redshifts (e.g. \citealt{2011ApJ...737...90B,
2012ApJ...759..135O,2012arXiv1212.5222M,2012ApJ...744..179S,2013MNRAS.429..150L,
2012ApJ...760..108B,2012ApJ...758...93F}).
Recently, measurements of the kinetic Sunyaev-Zel'dovich 
effect with the South Pole Telescope have been used to put limits on the epoch 
and duration of the reionization \citep{2012MNRAS.422.1403M,2012ApJ...756...65Z,2012arXiv1211.2832B}, 
though the obtained limits depend on the detailed 
physics of reionization \citep{2012ApJ...756...65Z,2013arXiv1301.3607P}.

The most promising probe of this evolutionary stage is the 21cm transition 
of neutral hydrogen (see \citealt{2006PhR...433..181F} for a review).
The EDGES\footnote{Experiment to Detect the Global EoR Signature, 
see http://www.haystack.mit.edu/ast/arrays/Edges/} experiment has put 
the first observational lower limit on the duration of the 
epoch of reionization (EoR) of $\Delta z > 0.06$ \citep{2010Natur.468..796B}, 
and using the GMRT\footnote{The Giant Metrewave Radio Telescope, 
see http://gmrt.ncra.tifr.res.in/}, \citet{2011MNRAS.413.1174P,2013arXiv1301.5906P} 
put upper limits on the neutral hydrogen power spectrum.
The upcoming low frequency interferometers such as 
LOFAR\footnote{The Low Frequency Array, see http://www.lofar.org/}, 
PAPER\footnote{The Precision Array for Probing the Epoch of Reionization, 
see http://eor.berkeley.edu/}, MWA\footnote{The Murchison Widefield Array, 
see http://www.mwatelescope.org/}, and 21CMA\footnote{The 21 Centimeter Array, 
see http://21cma.bao.ac.cn/} may be able to detect signatures of reionization, 
and the next generation instruments such as HERA\footnote{The Hydrogen Epoch of Reionization Array,
see http://reionization.org/} and 
SKA\footnote{The Square Kilometre Array, see http://www.skatelescope.org/}
may be able to map out the reionization process in more detail, and
reveal the properties of the first luminous objects.
Interpreting the upcoming data from these instruments requires
detailed modeling of the reionization process.

Motivated by the results of numerical simulations, 
\citet{2004ApJ...613....1F} developed a ``bubble model'' for the 
growth of HII regions during the early reionization era.  
In this model, at a given moment during the early stage of 
reionization, a region is assumed to be ionized if the total number of 
ionizing photons produced within exceeds the average number 
required to ionize all the hydrogen in the region,
otherwise it is assumed to be neutral, 
though there could be smaller HII regions within it. At the 
very beginning, the ionized regions are mostly the 
surroundings of the just-formed first stars or galaxies, 
but as the high density regions where first stars and galaxies formed are
strongly correlated, very soon these regions would 
grow larger and merge to contain several nearby galaxies. The bubble 
model treatment can deal with 
the fact that a region can be ionized by neighboring sources
rather than only interior galaxies.

In the bubble model the number of star 
forming halos and ionizing photons are calculated with the 
extended Press-Schechter model \citep{1991ApJ...379..440B,1993MNRAS.262..627L}. 
The criterion of ionization is equivalent to the condition that the
average density of the region exceeds a certain threshold value 
(ionization barrier). The mass function of the HII region can then 
be obtained from the excursion set model, i.e. by calculating 
the probability of a random walk trajectory first up-crossing the barrier. 
With a linear fit to the ionization barrier, \citet{2004ApJ...613....1F}
obtained the HII bubble mass function during the early stage of reionization
(see the next section for more details). This 
analytical model matches simulation results reasonably 
well \citep{2007ApJ...654...12Z}, and is much faster to compute than 
the radiative transfer numerical simulations, so it 
can be used to explore large parameter space. 
It also provides us an intuitive understanding on the physics of  
the reionization process. Instead of the full analytical calculation, 
one can also apply the same idea to make semi-numerical simulations 
\citep{2007ApJ...654...12Z,2007ApJ...669..663M,
2009ApJ...703L.167A,2009MNRAS.394..960C, 2012arXiv1212.6099Z}. 
In these simulations the 
density field is generated by the usual N-body simulation or the first order perturbation theory, 
the ionization field is then predicted with the same criteria as the analytical model.
The semi-numerical approach allows relatively 
fast computation, while at the same 
time providing three-dimensional visualization of the reionization process.

The bubble model also has certain limitations. As HII regions
form and grow, they begin to contact with each other,  
spherical ``bubbles'' are no longer a good 
description of the HII regions. After percolation of the HII regions, 
the photons 
from more distant regions, i.e. the ionizing background, 
become very important. Eventually the total volume fraction of the bubbles
predicted by the model would exceed one, and 
slightly before this moment the bubble model breaks down.
Although the bubble model may still be successful in some average
sense after percolation, and \citet{2007ApJ...654...12Z} indeed obtained fairly good
agreement between the model-based semi-numerical simulation and radiative transfer 
simulations even after ionized bubbles 
overlap, it is necessary to construct a more accurate
model for the late stage of reionization, to account for the non-bubble topology and 
the existence of an ionizing background.

One may consider to use similar reasonings to construct an 
analytical model for the remaining neutral regions after the percolation of
ionized regions. During this epoch, the high density of galaxies and minihalos 
allow them to have a higher recombination rate and thus remain neutral.
Besides these compact neutral regions, there are also 
large regions with 
relatively low density, which remain neutral because fewer galaxies 
formed within them. We shall call these neutral regions ``islands'', which 
remains above the flooding ionization for a moment. This is in some sense similar to the 
voids of large scale structure, just as the extended Press-Schechter model 
can predict the number of both halos and voids, we can also develop models of 
the neutral islands.  However, we do need to change the barrier to 
take into account the background ionizing photons in order to 
model the island evolution correctly.

On the observational aspect, the island distribution and its evolution 
are important for the 21cm signal, which directly relates to the 
neutral components in the Universe, and it would be relatively easier for the 
upcoming instruments to probe the signal at the late reionization 
stages, where the redshifted 21cm line have higher frequencies and weaker foregrounds.
Also, the neutral islands may also contribute to the overall opacity of the IGM in 
addition 
to the Lyman-limit systems, and in turn affect the evolution of the UV background
and the detectability of high redshift galaxies (e.g. \citealt{2013MNRAS.429.1695B}).

In this paper, we aim to construct an analytical island model, which is 
complementary to the bubble model. It applies to the neutral regions left 
over after the ionized bubbles overlap with each other, when the neutral 
islands are more isolated. Based on the excursion set formalism, we 
identify the islands by finding the first-crossings of the random 
walks {\it downward} the island barrier, which is deeper than the bubble barrier
because it takes into account the 
background ionizing photons in addition to the photons produced by stars
inside the island region. We then use the excursion set model to calculate the 
crossing probability at different mass scales, and derive the mass 
distribution function of the islands. 

However, inside the large neutral islands smaller ionized bubbles may also 
form. We investigate this ``bubbles-in-island'' problem by 
considering the conditional probability for the excursion trajectory 
to first down-cross the island barrier, then up-cross 
the original bubble barrier (without the contribution of the ionizing 
background) at a smaller scale. It turns out that a large number of bubbles 
may form inside the islands, such that a large fraction of the inside of some 
``islands'' is ionized. However, we may set 
a percolation threshold as an upper limit on the ``bubbles-in-island'' 
fraction, below which the islands are still relatively simple. We also try 
to shed light on the shrinking process of the islands, and obtain a 
coherent picture on the late stage of the epoch of reionization.

In the following, we first briefly review the excursion set theory and the bubble
model in \S\ref{reviewEST}, then we generalize it and develop the formalism
of  ``island model'' in \S\ref{Model}, and we employ a
simple toy model to illustrate the calculation. An important aspect of the theory 
is the treatment of the so called ``bubbles-in-island'' problem, i.e. self-ionized 
bubbles inside the neutral islands, we also discuss how to take this effect into account.
\S\ref{ion_back} presents our treatment of the ionizing background 
taking into account the absorption from Lyman-limit systems.
With these tools in hand we study the reionization process in \S\ref{results}, 
the consumption rate of background ionizing photons is assumed to be 
proportional to the surface area of the island. The size distribution of the 
islands are calculated for different redshifts. 
We summarize our results and conclude in \S\ref{Discuss}. 
Throughout this paper, we adopt the cosmological parameters from the
7-year {\it Wilkinson Microwave Anisotropy Probe} (WMAP7) measurements 
combined with BAO and $H_0$ data: $\Omega_b = 0.0455$, $\Omega_c = 0.227$, 
$\Omega_\Lambda = 0.728$, 
$H_{\rm 0} = 70.2\km\psec \Mpc^{-1}$, $\sigma_{\rm 8} = 0.807$ and 
$n_{\rm s} = 0.961$ \citep{2011ApJS..192...18K}, but the 
results are not sensitive to these parameters.

\section{A Brief Review of the Excursion Set Theory and the Bubble Model}\label{reviewEST}
\subsection{The Excursion Set Model}

Our island model is based on the excursion set theory. Here we give a 
brief review of the excursion set approach, especially its application 
to the reionization process, i.e. the bubble model. 
For a more comprehensive review 
of the excursion set theory and its extensions and applications, we refer the 
interested readers to \citet{2007IJMPD..16..763Z} and references therein.

\begin{figure}[t]
\centering{
\includegraphics[scale=0.4]{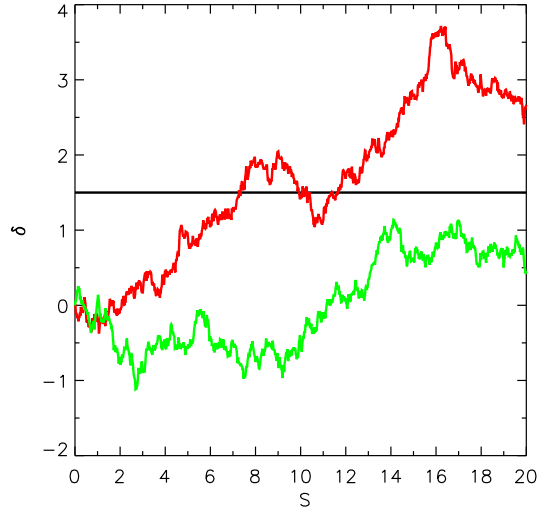}
\caption{Two random walk trajectories in the excursion 
set theory. Here $S=\sigma^2(M)$ denotes the variance 
of $\delta_{\rm M}$, which is the density fluctuation 
smoothed on mass scale $M$. All trajectories originates from
$(S,\delta)=(0,0)$. The horizontal line represents a flat 
barrier, motivated by spherical collapse.}
\label{Fig.trace}
}
\end{figure}

In what follows, we consider the density contrast field evaluated 
at some early time but extrapolated to the present 
day using linear perturbation theory. 
Consider a point $\mathbf{x}$ in space, 
the density contrast $\delta(\mathbf{x})$ around 
it depends on the smooth mass scale $M$ under consideration. The variance of 
the density fluctuations on scale $M$, $S=\sigma^2(M)$, monotonically 
decreases with increasing $M$ in our Universe, so we can 
use $S$ to represent the scale $M$. Starting at $M = \infty$, 
i.e. $S = 0$, we move to smaller and smaller 
scales surrounding the point of interest, and compute the smoothed 
density field as we go along. If we use a k-space top hat window function 
to smooth the density field, at each scale $k$ 
a set of independent Fourier modes 
are added, the trajectory of $\delta$ can be described by a random walk
where each step is independent, forming random
trajectories on the $S-\delta$ plane. Each of these trajectories starts
from the origin of the $(S,\delta)$ plane, with the variance of all trajectories given
by $\langle \delta^2 (S) \rangle = S$. Two sample trajectories are 
shown in Fig.~\ref{Fig.trace}. Typically, the trajectories jitter more and 
deviate farther from $\delta=0$ at larger $S$.

It is assumed that at redshift $z$ and on scale $M$, 
regions with average density above certain threshold value
$\delta_c$ will collapse into halos, while regions with average density below the threshold 
would remain uncollapsed. The galaxies formed inside sufficiently massive
halos. In some models, $\delta_c$ is only a function of redshift, more generally
it is a function of both redshift and mass scale. The formation of a halo 
corresponds to the trajectory up-crossing a barrier $\delta_c(M,z)$
in the $S-\delta$ plane. The excursion set theory was developed 
to compute the probabilities for such crossing, and then give the mass
distribution of the corresponding halos.

An important issue which must be addressed is the ``cloud-in-cloud'' problem. For a 
given central point, the critical threshold could be exceeded multiple times, 
corresponding to possible halos on different mass scales. In the excursion set
theory, one determines the largest smoothing scale $M$ (smallest $S$), 
at which a trajectory {\it first} up-crosses the halo barrier at $\delta_c$, and identify it as the 
halo at that redshift, while smaller scale crossings are ignored. Physically, it is reasonable to think 
that the smaller scale upcrossing corresponds to a small halo which formed earlier and merged into the 
bigger halo.

The probability of the barrier crossing can be computed by
solving a diffusion equation with the appropriate boundary conditions, and the {\it first crossing}
probability can be calculated with an absorbing barrier. For a constant density barrier 
and a starting point of $(\delta_0,S_0)$, the differential probability of first-crossing 
the barrier $\delta_c$ at $S$, known as the ``first-crossing distribution'', can be written as:
\begin{equation}
f(S|\delta_0,S_0) {\rm d}S = \frac{\delta_c-\delta_0}{\sqrt{2\pi}(S-S_0)^{3/2}}\, 
\exp \left[ - \, \frac{(\delta_c-\delta_0)^2}{2(S-S_0)}\right]\, {\rm d}S,
\end{equation}
and around the whole Universe, the mass function of the virialized halos is obtained by setting $S_0 = 0$ and $\delta_0 = 0$, which is
\begin{equation}\label{GeneralMF}
\frac{{\rm d}n}{{\rm d} \ln M} = \bar{\rho}_{\rm m,0} f(S) \left|\frac{{\rm d}S}{{\rm d}M}\right|.
\end{equation}

Besides the halo mass function, the excursion set theory can also be used to model the halo 
formation and growth \citep{1991ApJ...379..440B,1993MNRAS.262..627L}, and halo clustering 
properties \citep{1996MNRAS.282..347M}. Apart from the virialized halos, it could be applied to 
various structures in the Universe, such as the voids in the galaxy 
distribution \citep{2004MNRAS.350..517S,2012MNRAS.420.1648P,2006MNRAS.366..467F,2007MNRAS.382..860D} 
and the ionized bubbles during the 
early stages of reionization \citep{2004ApJ...613....1F}. It has also been extended to the case of
moving barriers \citep{2002MNRAS.329...61S,2006ApJ...641..641Z}. Strictly speaking, the 
probabilities given above is calculated for uncorrelated steps, which is correct for the 
k-space tophat filter but not for the real space tophat filter. The excursion set model with 
correlated steps have also been developed \citep{2008MNRAS.389..461P,2012MNRAS.420.1429P,
2012MNRAS.419..132P,2012MNRAS.423L.102M,2013arXiv1303.0337F,2013arXiv1306.0551M}, but below we will still use the uncorrelated model for its
simplicity.

\subsection{The Bubble Model}

In the excursion set model of ionized bubbles during reionization, 
i.e. the ``bubble model'', a region is considered ionized if it could emit sufficient
ionizing photons to get all the hydrogen atoms in the region ionized \citep{2004ApJ...613....1F}. 
Assuming that the number of the ionizing photons emitted is proportional to the total
collapse fraction of the region, the ionization condition can be written as 
\begin{equation}\label{Eq.bubbleCriterion}
f_{\rm coll} \ge \xi^{-1},
\end{equation}
where 
\begin{equation}
\xi = f_{\rm esc}\, f_\star\, N_{\rm \gamma/H}\, (1+\bar{n}_{\rm rec})^{-1}
\end{equation}
is  an ionizing efficiency factor,  in which $f_{\rm esc}$, $f_\star$, $N_{\rm \gamma/H}$, 
and $\bar{n}_{\rm rec}$ 
are the escape fraction, star formation efficiency, 
the number of ionizing photons emitted per H atom 
in stars, and the average number of recombinations 
per ionized hydrogen atom, respectively. 
For a Gaussian density field, the collapse fraction of a 
mass scale $M$ with the mean linear overdensity $\delta_{\rm M}$ at redshift $z$ can be 
written as \citep{1991ApJ...379..440B,1993MNRAS.262..627L}:
\begin{equation}\label{Eq.fcoll}
f_{\rm coll}(\delta_{\rm M}; M,z) = {\rm erfc} \left[
\frac{\delta_c(z)-\delta_{\rm M}}{\sqrt{2[S_{\rm max}-S(M)]}}\right],
\end{equation}
where $S_{\rm max} = \sigma^2(M_{\rm min})$,
in which $M_{\rm min}$ is the minimum collapse scale, and $\delta_c(z)$ 
is the critical density for 
collapse at redshift $z$ linearly extrapolated to the present time.
$M_{\rm min}$ is usually taken to be the mass corresponding to 
a virial temperature of $10^4 \K$, 
at which atomic hydrogen line cooling becomes efficient.
With this collapse fraction, the self-ionization constraint can be written as a barrier on 
the density contrast \citep{2004ApJ...613....1F}:
\begin{equation}\label{Eq.bubbleBarrier}
\delta_{\rm M} > \delta_{\rm B}(M,z) \equiv \delta_c(z) - \sqrt{2[S_{\rm
max} - S(M)]} \, {\rm erfc}^{-1} \left(\xi^{-1} \right).
\end{equation}
Solving for the {\it first}-up-crossing distribution of random walks with respect 
to this barrier, $f(S,z)$, 
the bubbles-in-bubble effect has been included, and
the size distribution of ionized bubbles can be obtained from Eq.(\ref{GeneralMF}),
then the average volume fraction of ionized regions can be written as:
\begin{equation}
Q^{\rm B}_{\rm V} = \int {\rm d}M \,\frac{{\rm d}n}{{\rm d}M}\, V(M). 
\end{equation}

In the linear approximate solution, $\delta_{\rm B}(M,z) = \delta_{\rm B,0} + \delta_{\rm B,1} S$, 
with the intercept on $S=0$ axis given by 
\begin{equation}
\delta_{\rm B,0} \equiv
\delta_c(z) - \sqrt{2\, S_{\rm max}}\, {\rm erfc}^{-1} \left(\xi^{-1} \right),
\label{eq:b0} 
\end{equation}
and the slope is 
\begin{equation}
\delta_{\rm B,1} \equiv \left. \frac{\partial \delta_{\rm B}}{\partial S} \right| 
_{S \rightarrow 0} = \frac{
  {\rm erfc}^{-1} \left(\xi^{-1} \right)}{\sqrt{2S_{\rm max}}}.
\label{eq:b1} 
\end{equation}
The number density of HII bubbles is then given by \citep{2004ApJ...613....1F}
\begin{equation}
M \frac{{\rm d}n}{{\rm d}M} = \frac{1}{\sqrt{2\,\pi}} \
\bar{\rho}_{\rm m,0} \ \left|\frac{{\rm d} S}{{\rm d} M} \right| \ 
\frac{\delta_{\rm B,0}}{S^{3/2}} \exp
  \left[ - \frac{\delta_{\rm B}^2(M,z)}{2\, S} \right].
 \end{equation}

According to the bubble model, at high redshifts the regions of high
overdensity were ionized earlier, because only in such regions 
galaxy-harboring halos formed, producing sufficient number of ionizing 
photons. In the excursion set theory, this is represented by those trajectories which 
excurse over the high barrier $\delta_{\rm B}(S)$.  As structures grow, the 
barrier function $\delta_{\rm B}(S)$ lowers, thus regions of 
relatively lower density become ionized. As the density 
and size of bubbles increase, they begin to
overlap. As long as the topology of the bubbles remains mostly discrete, this
description is valid. However, at a certain point, the intercept $\delta_{\rm B,0}$ drops low enough
to 0 that all trajectories which started out raising from the origin point of 
the $S-\delta$ plane would have crossed the barrier, 
and regions of the average density of the Universe would have been ionized. In fact,
the bubble description of HII regions perhaps failed slightly earlier, 
because when the ionized regions occupy a sizable fraction of the total volume, they become
connected, the topology becomes sponge-like, and it is no longer
possible to treat the ionized regions as individual bubbles.

\section{The Excursion Set Model of Neutral Islands}\label{Model}
\subsection{The general formalism}

The bubble model succeeds in describing the growth of HII regions 
before the percolation of HII regions.
As a natural generalization to the bubble model, we develop a model which 
is appropriate for the late stage of reionization, 
when the HII regions have overlapped with each other, 
and the neutral regions are more isolated and embedded in the sea of 
photon-ionized plasma and ionizing photons. According to the
bubble model, the regions with higher densities are ionized earlier, and by this 
stage even the regions of average density have been ionized, so the remaining large scale
neutral regions (``islands'') are underdense regions. Of course, besides these large
neutral regions, there are also galaxies and minihalos, in which neutral hydrogen exists
because they have very high density and hence high recombination rates, which keep them from
being ionized. We shall not discuss these small, highly dense HI systems in this paper, their 
number distribution can be predicted with the usual halo model formalism 
(see \citealt{2002PhR...372....1C} for a review).
The neutral islands during the late era of reionization are more likely isolated than the ionized bubbles, 
similar to the voids at lower redshifts.

In the island model, we assume that most part of the Universe has been ionized, 
but the reionization has not been completed. 
The condition for a region remains neutral is just the opposite of the ionization 
condition, that is, the total number of ionizing photons 
is less than the number required to ionize all hydrogen atoms in the region.
At this stage, however, it is also important to include the background ionizing 
photons which are produced outside the region.
An island of mass scale $M$ at redshift $z$ has to 
satisfy the following condition in order to remain neutral:
\begin{equation}\label{Eq.IslandCondition}
\xi f_{\rm coll}(\delta_{\rm M}; M,z)+ \frac{\Omega_m}{\Omega_b} \frac{N_{\rm back} m_{\rm H} 
}
{M X_{\rm H} (1+\bar{n}_{\rm rec})} < 1,
\end{equation}
where $N_{\rm back}$ is the number of background ionizing photons that are 
consumed by the island, and $X_{\rm H}$ is the mass fraction of the baryons in hydrogen.
The first term on the L.H.S. is due to self-ionization, while the second term is 
due to the ionizing background. Note that in the usual convention of the bubble model,
the  number of recombination factor $(1+\bar{n}_{\rm rec})^{-1}$ is absorbed in the $\xi$
parameter, and to be consistent with these literatures here we follow this convention, but 
we should keep in mind that if one changes $\bar{n}_{\rm rec}$, the adopted $\xi$ value
should be changed accordingly.

Using Eq.~(\ref{Eq.fcoll}), 
the condition (\ref{Eq.IslandCondition}) can be rewritten as a
constraint on the overdensity of the region:
\begin{equation}\label{Eq.islandBarrier}
\delta_{\rm M} < \delta_{\rm I}(M,z) \equiv \delta_c(z) - \sqrt{2[S_{\rm
max} - S(M)]} \, {\rm erfc}^{-1} \left[K(M,z)\right],
\end{equation}
where
\begin{equation}\label{Eq.K}
K(M,z) = \xi^{-1} \left[1 - N_{\rm back} (1+\bar{n}_{\rm rec})^{-1}
\frac{m_{\rm H}} {M (\Omega_b / \Omega_m) X_{\rm H}}\right].
\end{equation}
Due to the contribution of the ionizing background photons, in the excursion set model
the barrier for the neutral islands is different from the barrier used in 
the bubble model (Eq.~(\ref{Eq.bubbleBarrier})), as the ionizing background would not be present  
when the bubbles are isolated.  Below, we shall call a barrier with only the
self-ionization term the ``bubble barrier'', denoted by $\delta_{\rm B}(M,z)$, since 
it is used to compute the probability of forming bubbles. 
Inclusion of the ionizing background would make the barrier 
much more negative, and we shall  
call the full barrier the ``island barrier'', denoted by $\delta_{\rm I}(M,z)$.

As discussed in the last section, the bubble 
barrier lowers as the structure formation 
progresses. Even if we simply compute the barrier as in the original bubble model, 
i.e. including only the ionizing photons from collapsed halos within the 
region being considered, it could have negative intercepts, i.e. $\delta_{\rm B}(S=0)<0$ 
(see e.g. the thin lines in Fig.~\ref{Fig.barrierV}).
When bubble barrier passes through the origin point of the $\delta - S$ plane, all regions with 
the mean density $\delta=0$ are ionized, this means that most of the Universe is ionized.
It is also from this moment onward a global ionizing background is gradually set up. 
We will define the redshift when this occurred as 
the ``background onset redshift'' $z_{\rm back}$, 
and it can be solved from the following equation:
\begin{equation}
\delta_{\rm I}(S=0;z=z_{\rm back}) = \delta_c(z_{\rm back}) - \sqrt{2\, S_{\rm max}(z_{\rm back})}
\;{\rm erfc}^{-1}(\xi^{-1}) = 0.
\end{equation}
We take $\{f_{\rm esc}, f_{\star}, N_{\rm \gamma/H}, \bar{n}_{\rm rec}\}
=\{0.2, 0.1, 4000, 1\}$ as the fiducial set of parameters, 
so that $\xi=40$ and $z_{\rm back}=8.6$, consistent with the 
observations of the quasars/gamma-ray bursts absorption 
spectra \citep{2008MNRAS.386..359G,2008MNRAS.388L..84G} and Lyman alpha emitters 
surveys (e.g. \citealt{2006ApJ...647L..95M,2007ApJ...671.1227D}) which 
suggests $x_{\rm HI} \ll 1$ at $z \approx 6$. 
We note that this background onset redshift is also consistent with our ionizing background
model presented in \S\ref{ion_back}, in which the intensity of the ionizing background
starts to rapidly increase around redshift $z \sim 8 - 9$ (see Fig.~\ref{Fig.Gamma_12}).
However, the exact value of this background onset redshift has little impact on
the final model predictions on the island distribution, as the ionizing background increases
quite rapidly during the late stage of reionization (see \S\ref{ion_back}) and 
the main background contribution to the ionizations comes from the redshift range
just above the redshift under consideration.

As all trajectories start from the point $(S,\delta)=(0,0)$, and the island barrier has a negative intercept,
we see that instead of the usual up-crossing condition 
in the excursion set model, here the condition of forming a neutral island is represented
by a {\it down-crossing} of the barrier. Once a random walk trajectory hits 
the island barrier, we identify an island with the crossing scale, 
and assign the points inside this region to a neutral island of the appropriate mass.
Similar to the ``cloud-in-cloud'' problem in the halo 
model \citep{1991ApJ...379..440B}, 
or the ``void-in-void'' problem in the void model \citep{2004MNRAS.350..517S}, 
there is also an ``island-in-island'' problem. As in those cases, this problem can 
also be solved naturally by considering only the 
{\it first}-{\it down}-crossings of the barrier curve.

For a general barrier, \citet{2006ApJ...641..641Z} developed an intergral equation method
for computing the first-{\it up}-crossing distribution. 
Similarly, denoting the island scale with its variance $S_{\rm I}$, the first-{\it down}-crossing 
distribution of random trajectories with an arbitrary island barrier can be solved as:
\begin{equation}
f_{\rm I}(S_{\rm I}) = -g_1(S_{\rm I}) - \int_0^{S_{\rm I}} {\rm d}S'
f_{\rm I}(S')\left[g_2(S_{\rm I},S')\right],
\end{equation}
where
\begin{equation}
g_1(S_{\rm I}) = \left[\frac{\delta_{\rm I}(S_{\rm I})}{S_{\rm I}}
-2\frac{{\rm d}\delta_{\rm I}}{{\rm d}S_{\rm I}}\right] P_0[\delta_{\rm I}(S_{\rm I}),S_{\rm I}],
\end{equation}
\begin{equation}
g_2(S_{\rm I},S') = \left[2\frac{{\rm d}\delta_{\rm I}}{{\rm d}S_{\rm I}} 
- \frac{\delta_{\rm I}(S_{\rm I})-\delta_{\rm I}(S')}{S_{\rm I}-S'}\right]
P_0[\delta_{\rm I}(S_{\rm I})-\delta_{\rm I}(S'),S_{\rm I}-S'],
\end{equation}
and $P_0(\delta,S)$ is the normal Gaussian distribution with variance $S$, which is defined as
\begin{equation}
P_0(\delta,S) = \frac{1}{\sqrt{2\pi S}} \exp
\left(-\frac{\delta^2}{2S}\right).
\end{equation}
These integral equations can be solved numerically with the algorithm of 
\citet{2006ApJ...641..641Z}, we can then obtain
the mass function of islands at redshift $z$:
\begin{equation}
\frac{{\rm d}n}{{\rm d}\ln M_{\rm I}}(M_{\rm I},z) 
= \bar{\rho}_{\rm m,0} f_{\rm I}(S_{\rm I},z) \left|\frac{{\rm d}S_{\rm I}}{{\rm d}M_{\rm I}}\right|.
\label{eq.hostMF}
\end{equation}
With the neutral island mass function, the volume fraction of neutral regions is given by
\begin{equation}
Q^{\rm I}_{\rm V} = \int {\rm d}M_{\rm I}\,\frac{{\rm d}n}{{\rm d}M_{\rm I}}\, V(M_{\rm I}). 
\end{equation}


\subsection{A toy model with island-permeating ionizing background photons}
\label{toy_model}

To illustrate the basic ideas of the island model, let us 
consider a toy model in which the ionizing photons permeated through the 
neutral islands with a uniform density. This is not a physically realistic model, because if
ionizing photons can permeate through the neutral regions with sufficient flux, 
there would be no distinct ionizing bubbles or
neutral islands, though it may be possible to have a small component of 
penetrating radiation such as hard X-rays, but that would be much smaller than the 
total ionizing background. 
The reason we consider this model is that it is possible to derive 
a simple analytical solution, which could illustrate some aspects of the island model.

The island-permeating ionizing background photons are likely to be hard X-rays, 
whose mean free paths are extremely large even in the IGM with a high neutral fraction.
Therefore, here we use an extremely simple model for the ionizing background, 
in which the absorptions by dense clumps are neglected, and the mean free path of 
these background photons are comparable with the Hubble scale. In any case, this is a toy
model, a more realistic model for the ionizing background will be described in the next section.
Further, we assume that the total number of ionizing photons produced by
redshift $z$ is proportional to the total collapse fraction of 
the Universe at that redshift. Some of these photons would have already been consumed by 
ionizations took place before that redshift, and the ionizing background photons are 
what left behind. The comoving number density of background ionizing photons is then given by
\begin{equation}
n_\gamma = \bar{n}_{\rm H}\, f_{\rm coll}(z)\,
f_\star\, N_{\rm \gamma/H}\, f_{\rm esc} - (1-Q^{\rm I}_{\rm V})\,\bar{n}_{\rm H}\,(1+\bar{n}_{\rm rec}),
\label{eq.n_gamma}
\end{equation}
where $\bar{n}_{\rm H}$ is the average comoving number density of hydrogen in the Universe, 
and the other parameters are the same as those in Eq.(\ref{Eq.bubbleCriterion}).
The number density of ionizing photons given by Eq.~(\ref{eq.n_gamma}) depends on
the global neutral fraction $Q^{\rm I}_{\rm V}$, which is only known after we have applied the 
ionizing background intensity itself and solved the reionization model, so this equation 
should be solved iteratively.

\begin{figure}[t]
\centering{
\includegraphics[scale=0.4]{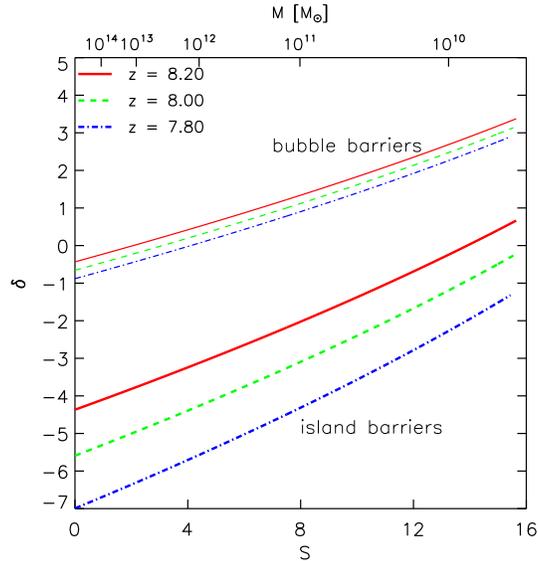}
\caption{The island barriers in the model with uniform island-permeating ionizing background
photons, the barriers are plotted for redshifts 8.2, 8.0 and 7.8 as thick curves from top to bottom respectively.
Here we assume $\{f_{\rm esc}, f_{\star}, N_{\gamma/H}, 
\bar{n}_{\rm rec}\}=\{0.2, 0.1, 4000, 1\}$. 
The bubble barriers (without ionizing background) at the same set of redshifts 
are shown as thin curves. On the top of figure box we also show the mass scales corresponding to $S$ for
reference.} 
\label{Fig.barrierV}
}
\end{figure}

Suppose that the background ionizing photons are uniformly distributed and consumed 
within the islands, then $N_{\rm back}$ is proportional to the island volume. 
We see from Eq.(\ref{Eq.K}) that $N_{\rm back}$ cancels
with the island mass $M$ in the denominator, and
we have $N_{\rm back}/M = n_\gamma/ \bar{\rho}_{\rm m}$. Therefore,
in this model, the $K$ factor is essentially independent of $M$, i.e. $K(M,z) = K(z)$,
then the island barrier becomes:
\begin{equation}
\delta_{\rm I}(M,z) = \delta_c(z) - \sqrt{2\, [S_{\rm
max} - S(M)]} \, {\rm erfc}^{-1} \left[K(z)\right].
\end{equation}
For a given redshift, $K=$constant, so similar to the bubble barrier, 
the only dependence of the island barrier on mass scale comes 
from $S(M)$. Taking the fiducial set of parameters, 
we plot the island barriers at redshift 8.2, 8.0 
and 7.8 in Fig.~\ref{Fig.barrierV} with thick curves from top to bottom respectively. The
bubble barriers are also plotted with thin lines in the same figure. 
Indeed, in this case the island barriers have the similar shape as the bubble barriers.
Both barriers increase with $S$, as shown in Fig.~\ref{Fig.barrierV}.

\begin{figure}[t]
\centering{
\includegraphics[scale=0.4]{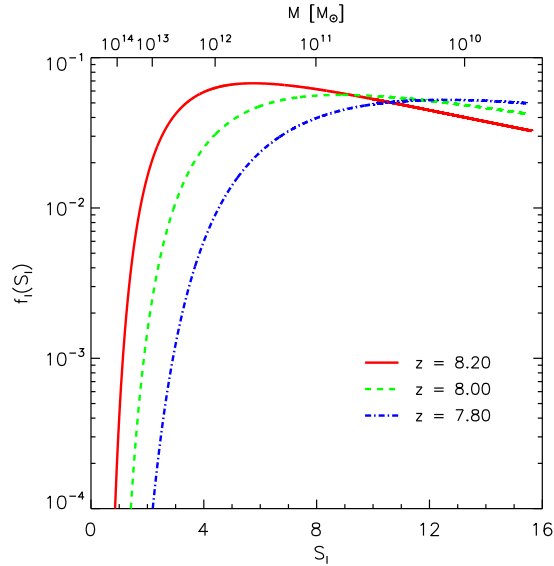}}
\caption{The first-down-crossing distribution in the island-permeating photon model
as a function of the island scale at redshifts 8.2, 8.0 and 7.8 from top to bottom
respectively.} 
\label{Fig.fI_V}
\end{figure}

As the redshift decreases, the linearly extrapolated critical overdensity  
$\delta_c(z)$ decreases, and both barriers move downward. 
For a given set of parameters, as the redshift decreases, $n_\gamma$ 
increases and $\bar{\rho}_m$ decreases, so that $N_{\rm back}/M$ increases. As a result, 
the island barrier decreases faster than the bubble barrier for the same decrease in redshift.
We cut all the curves in the figure at $\xi\,M_{\rm min}$, which is the scale 
for which a halo of $M_{\rm min}$ can ionize, and this set the lower limit of 
a bubble. 
In this toy model, we also cut the island scale at $\xi\,M_{\rm min}$, because
at smaller scales, the non-linear effect becomes important, and the collapse fraction
computed from the extended Press-Schechter model (Eq.(\ref{Eq.fcoll})), which is valid for Gaussian 
density field, is not accurate anymore. The exact value of the cutoff mass is not critical 
for the illustrative purpose here. Note that this mass-cut of islands is not necessary
for the more realistic island model presented in \S\ref{results}, in which the lower limit of 
an island scale is naturally set by the survival limit of islands in the presence of an ionizing background 
(see the text in \S\ref{results}).
Below this scale, the neutral hydrogen exists only in minihalos or galaxies. 

The first-down-crossing distribution for the islands in the 
island-permeating photons model 
is plotted for three redshifts in Fig.~\ref{Fig.fI_V}, $S$ and corresponding mass scale $M$
are shown on the bottom and top axes respectively. As expected, at small $S$  
the down-crossing probability is vanishingly small, because in this region the barrier 
is very negative, and the average displacement of the random trajectories is still very 
small. As $S$ increases, the trajectories excurse with wider ranges, and 
in this model the barriers also raise up with increasing $S$, so the crossing probability
increases rapidly. 
For $z=8.2$, the probability peaks at $S_{\rm I} \approx 5.8$ 
with $f_{\rm I} \approx 0.07$, 
then begins to decrease, because for many trajectories the first crossing happened earlier. 
As the redshift decreases,  
the island barrier moves downward rapidly, and it becomes harder and harder to down-cross it at large scales,
with most of the first down-crossings happen at smaller scales.   
As a result, the first-down-crossing probability decreases very rapidly at large scale, 
and it increases at small scales.

\begin{figure}[t]
\centering{
\subfigure{\includegraphics[scale=0.4]{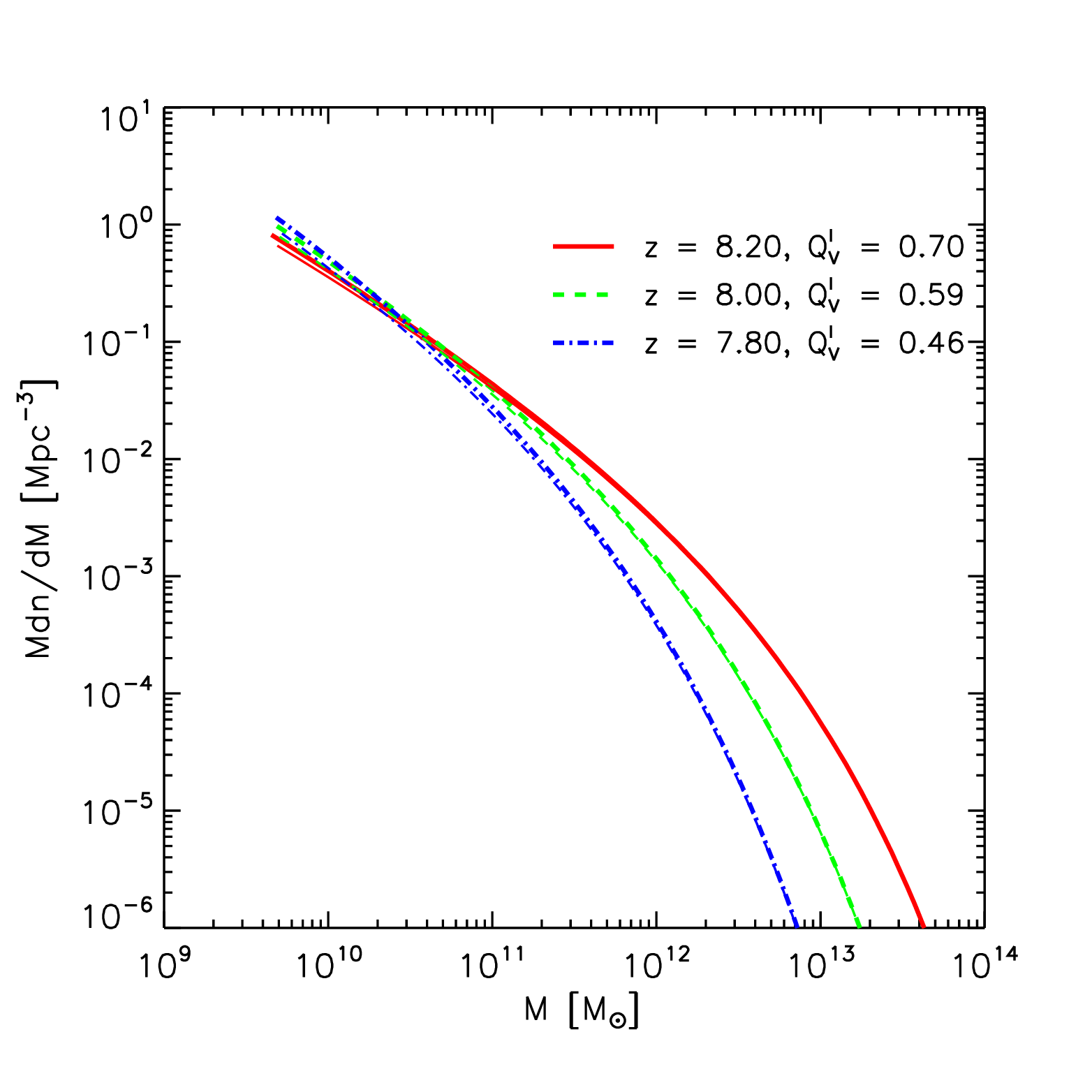}}
\subfigure{\includegraphics[scale=0.4]{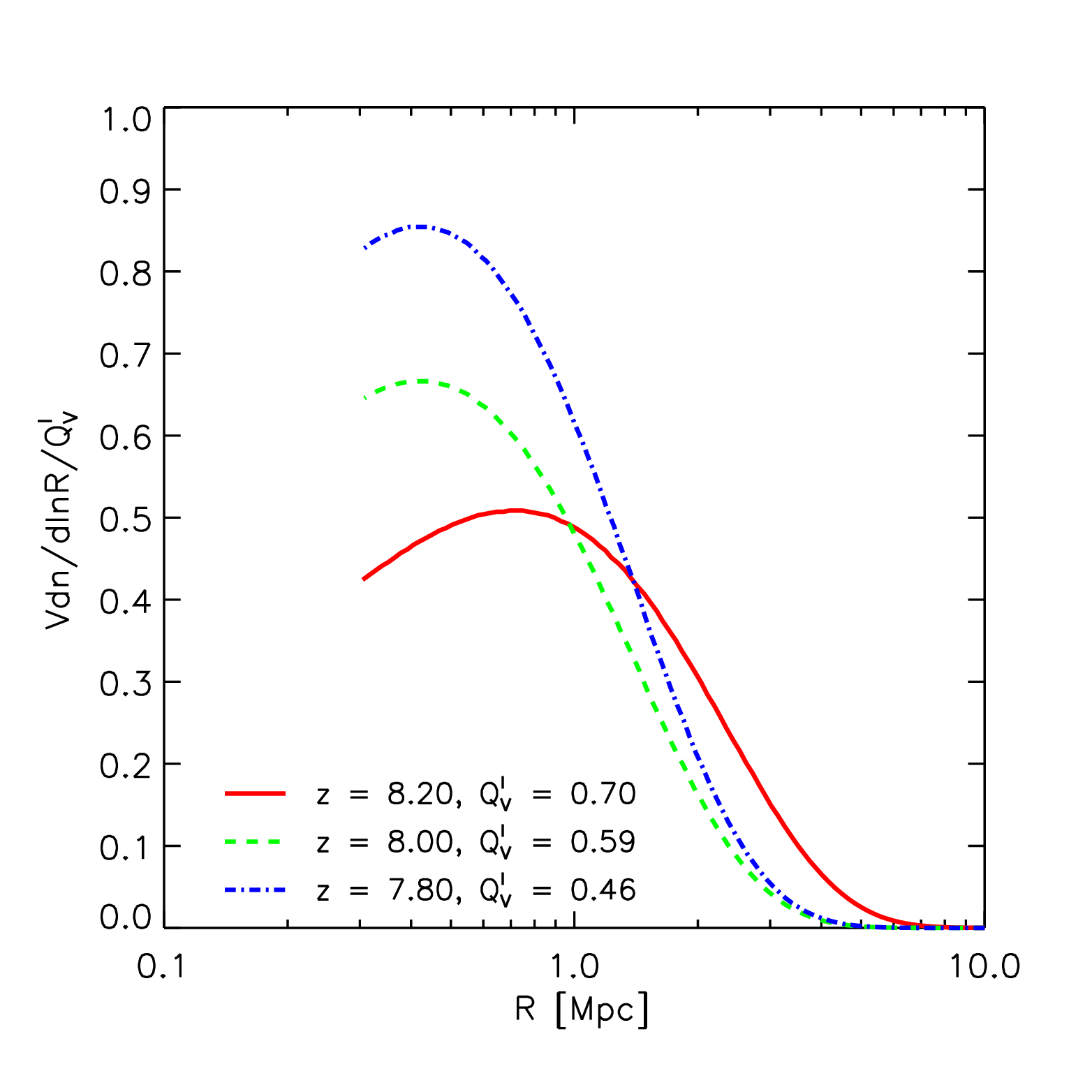}}   
\caption{{\it Left panel}: The number distribution functions of neutral islands in the model with 
a uniform island permeating ionizing background. The numerical solutions are shown as thick curves 
for redshifts 8.2, 8.0 and 7.8 from top to bottom on the right respectively,  
the corresponding volume filling factors of islands are $Q^{\rm I}_{\rm V} =$ $0.70\; (z = 8.2)$, $0.59\; (z = 8.0)$, 
and $0.46\;  (z = 7.8)$, respectively. The thin curves show the distribution function given by the 
analytical form in the linear approximation.
{\it Right panel}: The size distributions of islands at the 
same redshifts as in the left panel, normalized by the total neutral fraction $Q^{\rm I}_{\rm V}$.}
\label{Fig.MF_Rdistr_V}
}
\end{figure}

The mass functions of islands at three redshifts are plotted in the left panel 
of Fig.~\ref{Fig.MF_Rdistr_V}. The volume filling factors of the neutral islands are 
$Q^{\rm I}_{\rm V} =0.70\; (z = 8.2)$, $0.59\; (z = 8.0)$, 
and $0.46\;  (z = 7.8)$, respectively, and the corresponding ionizing background
can be expressed as an HI photoionization rate of $\Gamma_{\rm HI} = n_\gamma (1+z)^3 \,c\,\sigma_i
\approx 1.6\times10^{-11} \psec$.
Here $\sigma_i$ is the frequency averaged photoionization cross-section 
of hydrogen.
This level of the ionizing background is unreasonably high, because in this toy 
model, we have neglected the effects of dense clumps, minihalos, 
and any other possible absorbing systems that could limit
the mean free path of the ionizing photons.
To facilitate comparisons with the bubble 
distribution function in \citet{2004ApJ...613....1F}, we also plot in the right 
panel the volume weighted distribution of the effective radii of the islands 
computed assuming that the islands 
are uniform spheres, normalized by the total neutral fraction as in the bubble model.
Note that 
\begin{equation}\label{Eq.R}
V \frac{{\rm d}n}{{\rm d}\ln R} \propto 3M^2 \frac{ {\rm d}n}{{\rm d}M} \propto M\frac{ {\rm d}n}{{\rm d}\ln M},
\end{equation}
so this also reflects how masses are distributed in islands of different sizes. 

Unsurprisingly, within a given volume, small size islands 
are much more numerous than larger ones, as shown in the left panel. 
Similar to the general shape of the volume weighted bubble size distribution in the bubble model, 
there is a peak in the island size distribution at each redshift in this model. This means that 
in the photon-permeating model, the neutral mass is dominated by those islands with the characteristic
scale where the distribution peak locates.
As redshift decreases, the left panel of Fig.~\ref{Fig.MF_Rdistr_V}
shows that the number of large islands decreases rapidly, while the number of the
smallest ones even increases a little. This evolutionary behavior is also shown in the right panel
of Fig.~\ref{Fig.MF_Rdistr_V}, 
in which large bubbles gradually disappeared, resulting in a raising curve on the 
small $R$ end.

In fact, for this toy model, the barrier shape is very close to a straight line, for which 
simple analytical solution exists and is very accurate. If we expand the barrier as a linear function of 
$S$, we have 
\begin{equation}
\delta_{\rm I} (M,z) = \delta_{\rm I,0}+\delta_{\rm I,1}\,S,
\end{equation}
where the intercept is 
\begin{equation}
\delta_{\rm I,0} \equiv \delta_c(z) - \sqrt{2\, S_{\rm max}}\, {\rm erfc}^{-1} \left[K(z) \right],
\label{eq:Ib0} 
\end{equation}
and the slope is
\begin{equation}
\delta_{\rm I,1} \equiv \frac{{\rm erfc}^{-1} \left[K(z) \right]}{\sqrt{2\,S_{\rm max}}}.
\end{equation}
Then the mass function of the host islands can be expressed analytically:
\begin{equation}\label{Eq.AnalyticMF}
M_{\rm I} \frac{{\rm d}n}{{\rm d}M_{\rm I}} = \frac{1}{\sqrt{2\,\pi}} \
\bar{\rho}_{\rm m,0} \ \left|\frac{{\rm d} S}{{\rm d} M_{\rm I}} \right| \ 
\frac{|\delta_{\rm I,0}|}{S^{3/2}(M_{\rm I})} \exp
  \left[ - \frac{\delta_{\rm I}^2(M_{\rm I},z)}{2\, S(M_{\rm I})} \right].
\end{equation}
These are plotted as thin lines in the left panel of Fig.~\ref{Fig.MF_Rdistr_V}, 
we see they almost coincide with the results of numerical solutions (thick lines).

The model of this subsection is only for demonstrating the formalism of calculation with 
additional (background) ionizing photons, and for simplicity we assumed that the 
consumed photons are proportional to the island volume. 
This is not realistic, because the ionization 
caused by a background is more likely proportional to the surface area $\Sigma$ of the island.
In the next sections we shall consider more realistic models.

\subsection{The Bubbles In Islands}
\label{bubbles_in_island}

Before moving to more realistic models, let us address the problem 
of ``bubbles-in-island'' first. 
In the above we have assumed that the neutral islands are simple spherical regions, but 
in fact there might also be self-ionized regions inside an island. 
This ``bubbles-in-island'' problem is similar but in the opposite sense of 
the ``voids-in-cloud'' problem in the void 
model \citep{2004MNRAS.350..517S,2012MNRAS.420.1648P}. 

We identify the bubbles inside neutral islands in the excursion set framework
by considering the trajectories which first 
down-crossed the island barrier $\delta_{\rm I}$ at $S_{\rm I}$, then at a larger $S_{\rm B}$ up-crossed over  
the bubble barrier $\delta_{\rm B}$. The bubble barrier is the barrier 
defined without considering the ionizing background, since this 
background should be absent inside large neutral regions. Note that in the 
toy model discussed above, the ionizing background permeates through the neutral 
islands, then it does not make sense to distinguish the island barriers outside and 
the bubble barriers inside, and
the problem of bubbles-in-island can not be discussed.

In the following, we denote the {\it host} island scale (including the bubbles inside)
and the bubble scale by $S_{\rm I}$ and $S_{\rm B}$ respectively, the first 
down-crossing distribution by $f_{\rm I}(S_{\rm I},\delta_{\rm I})$, and denote the 
conditional probablity for a bubble form inside as   
$f_{\rm B}(S_{\rm B},\delta_{\rm B}|S_{\rm I},\delta_{\rm I})$.
The probability distribution of finding a bubble of size $S_{\rm B}$ in a host island of 
size $S_{\rm I}$ is then given by
\begin{equation}
\mathcal{F}(S_{\rm B},S_{\rm I})=f_{\rm I}(S_{\rm I},\delta_{\rm I})~
\cdot ~f_{\rm B}(S_{\rm B},\delta_{\rm B}|S_{\rm I},\delta_{\rm I}).
\end{equation}
The neutral mass of an island is given by the total mass of the host
island minus the masses of bubbles of various sizes embedded in the host island, i.e.
\begin{equation}
M = M_{\rm I}(S_{\rm I}) - \sum_i M_{\rm B}^{i} (S_{\rm B}^{i}).
\end{equation}
The conditional probability distribution 
$f_{\rm B}(S_{\rm B},\delta_{\rm B}|S_{\rm I},\delta_{\rm I})$ characterizes the size 
distribution of bubbles inside an island of scale $S_{\rm I}$ and 
overdensity $\delta_{\rm I}$, and 
$f_{\rm B}(S_{\rm B},\delta_{\rm B}|S_{\rm I},\delta_{\rm I}) {\rm d}S_{\rm B}$ is the 
conditional probability of a random walk which first up-crosses $\delta_{\rm B}$ 
at between $S_{\rm B}$ and $S_{\rm B}+dS_{\rm B}$ given a starting point 
of $(S_{\rm I},\delta_{\rm I})$.

In order to compute $f_{\rm B}$, we could effectively shift the origin point
of coordinates to the point $(S_{\rm I},\delta_{\rm I})$, then the 
method developed by \citet{2006ApJ...641..641Z} is still 
applicable. The effective bubble barrier becomes:
\begin{equation}
\delta_{\rm B}^{\prime} = \delta_{\rm B}(S+S_{\rm I}) - \delta_{\rm I}(S_{\rm I}),
\end{equation} 
where $S = S_{\rm B} - S_{\rm I}$.
Given an island $(S_{\rm I},\delta_{\rm I})$, on average, the fraction of volume (or mass) 
of the island occupied by bubbles of different sizes is
\begin{equation}
q_{\rm B}(S_{\rm I},\delta_{\rm I};z) = \int_{S_{\rm I}}^{S_{\rm max}(\xi\cdot M_{\rm min})} 
\left[1+\delta_{\rm I}\,D(z)\right]~f_{\rm B}(S_{\rm B},
\delta_{\rm B}|S_{\rm I},\delta_{\rm I})~ {\rm d}S_{\rm B}.
\end{equation}
The factor $[1+\delta_{\rm I}\,D(z)]$ enters because these bubbles 
are in the environment with underdensity of $\delta_{\rm I}\,D(z)$, where $D(z)$ is 
the linear growth factor.
Then the net neutral mass of the host island can be written as 
$M=M_{\rm I}(S_{\rm I}) \,[1-q_{\rm B}(S_{\rm I},\delta_{\rm I};z)]$.
Taking into account the effect of bubbles-in-island, the neutral mass function 
of the islands at redshift $z$ is
\begin{equation}
\frac{{\rm d}n}{{\rm d}M}(M,z) 
= \frac{{\rm d}n}{{\rm d}M_{\rm I}} \frac{{\rm d}M_{\rm I}}{{\rm d}M} 
 = \frac{\bar{\rho}_{\rm m,0}}{M_{\rm I}} f_{\rm I}(S_{\rm I},z) 
\left|\frac{{\rm d}S_{\rm I}}{{\rm d}M_{\rm I}}\right| \frac{{\rm d}M_{\rm I}}{{\rm d}M}.
\end{equation}

\section{The ionizing background}\label{ion_back}

The intensity of the ionizing background is very important in the late reionization epoch. 
However, it has only been constrained after reionization from the mean transmitted flux 
in the Ly-$\alpha$ forest (e.g. \citealt{2011MNRAS.412.1926W,2011MNRAS.412.2543C}), and in 
any case it evolves with redshift and depends on the detailed history of the reionization.
Conversely, the evolution of the ionizing background also affects the reionization process.

In the toy model presented in \S\ref{toy_model}, we considered an island-permeating ionizing
background, for which the absorptions from dense clumps are neglected, and the resulting 
intensity of the ionizing background is unreasonably high.
Here we give a more realistic model for the ionizing background. Due to the existence of dense 
clumps that have high recombination rate and limit the mean free path of the ionizing 
background photons, an island does not see all the ionizing photons emitted by all the sources, 
but only out to a distance of roughly the mean free path of the ionizing photons.
The comoving number density of background ionizing photons at redshift $z$ can be modeled as
the integration of escaped ionizing photons that are emitted from newly collapsed objects 
and survived to the distances between the sources and the position under consideration:
\begin{equation}\label{n_gamma}
n_\gamma(z) \;=\; \int_z\, \bar{n}_{\rm H}\, \left|\frac{{\rm d}f_{\rm coll}(z')}{{\rm d}z'}\right|\, f_\star\, N_{\rm \gamma/H}\, f_{\rm esc}\, \exp \left[\,-\, \frac{l(z,z')}{\lambda_{\rm mfp}(z)}\right]\, {\rm d}z',
\end{equation}
where $l(z,z')$ is the physical distance between the source at redshift $z'$ and the redshift $z$ under 
consideration, and $\lambda_{\rm mfp}$ is the physical mean free path of the background 
ionizing photons.

Various absorption systems could limit the mean free path of the background ionizing photons.
The most frequently discussed absorbers are Lyman limit systems, which have large enough HI column density to keep self-shielded (e.g. \citealt{2000ApJ...530....1M,2005MNRAS.363.1031F,2013MNRAS.429.1695B}).
Minihalos are also self-shielding systems that could block ionizing photons. \citet{2005MNRAS.363.1031F}
developed a simple model for the mean free path of ionizing photons in a Universe where minihalos
dominate the recombination rate. However, as also mentioned in \citet{2005MNRAS.363.1031F}, the formation and the abundance of minihalos are highly uncertain \citep{2003MNRAS.346..456O}, and minihalos would be probably evaporated during the late epoch of reionization \citep{1999ApJ...523...54B,2004MNRAS.348..753S}, although they may consume substantial ionizing photons before they are totally evaporated \citep{2005MNRAS.361..405I}.
In addition to Lyman limit systems and minihalos, the accumulative absorption by low column density systems can not be neglected \citep{2005MNRAS.363.1031F}, but the quantitative contribution from these systems are quite uncertain, and need to be calibrated by high resolution simulations or observations.

Here we focus on the effect of Lyman limit systems on the mean free path of ionizing photons, and use a simple model for the IGM density distribution developed by \citet{2000ApJ...530....1M} (hereafter MHR00).
In the MHR00 model, the volume-weighted density distribution of the IGM measured from 
numerical simulations can be fitted by the formula
\begin{equation}
P_{\rm V}(\Delta)\, {\rm d}\Delta \;=\; A_0\, \exp \left[\, -\, \frac{(\Delta^{-2/3} - C_0)^2}{2\,(2\delta_0/3)^2}\, \right]\, \Delta^{-\beta}\, {\rm d}\Delta
\end{equation}
for $z\sim 2 - 6$, where $\Delta = \rho/\bar{\rho}$. Here $\delta_0$ and $\beta$ are parameters fitted 
to simulations. The value of $\delta_0$ can be extrapolated to higher redshifts by the function
$\delta_0 \,=\, 7.61/(1+z)$ \citep{2000ApJ...530....1M}, and we take $\beta = 2.5$ for the redshifts
of interest. The parameters $A_0$ and $C_0$ are set by normalizing $P_{\rm V}(\Delta)$ and 
$\Delta P_{\rm V}(\Delta)$ to unity.

Using the density distribution of the IGM, the mean free path of ionizing photons can be determined
 by the mean distance between self-shielding systems with relative densities above a critical value $\Delta_{\rm crit}$, and can be written as \citep{2000ApJ...530....1M,2005MNRAS.361..577C}
\begin{equation}\label{Eq.mfp}
\lambda_{\rm mfp} \;=\; \frac{\lambda_0}{[1\,-\, F_{\rm V}(\Delta_{\rm crit})]^{2/3}},
\end{equation}
where $F_{\rm V}(\Delta_{\rm crit})$ is the volume fraction of the IGM occupied by regions
with the relative density lower than $\Delta_{\rm crit}$, given by
\begin{equation}
F_{\rm V}(\Delta_{\rm crit}) \;=\; \int_0^{\Delta_{\rm crit}} P_{\rm V}(\Delta) \,{\rm d}\Delta.
\end{equation}
Following \citet{2001ApJ...559..507S}, and assuming photoionization equilibrium and case A recombination rate, the critical relative density for a clump to self-shield can be approximately 
written as (see also \citealt{2000ApJ...530....1M,2005MNRAS.363.1031F,2013MNRAS.429.1695B}):
\begin{equation}\label{critical_density}
\Delta_{\rm crit} \;=\; 36\, \Gamma_{-12}^{2/3}\, T_4^{2/15}\, \left(\frac{\mu}{0.61}\right)^{1/3}\,
\left(\frac{f_e}{1.08}\right)^{-2/3}\, \left(\frac{1+z}{8}\right)^{-3},
\end{equation}
where $\Gamma_{-12}\,=\, \Gamma_{\rm HI}/10^{-12} \psec$ is the hydrogen photoionization rate
in units of $10^{-12} \psec$, $T_4 \,=\, T/10^4 \K$ is the gas temperature in units of $10^4 \K$,
$\mu$ is the mean molecular weight, and $f_e \,=\, n_e/n_{\rm H}$ is the free electron fraction with respect to hydrogen. For the mostly ionized IGM during the late stage of reionization, we assume $T_4 = 2$.

The HI photoionization rate $\Gamma_{\rm HI}$ in Eq.(\ref{critical_density}) is related to the total
 number density of ionizing photons $n_\gamma$ in Eq.(\ref{n_gamma}) by
 \begin{equation}
 \Gamma_{\rm HI} \;=\; \int \, \frac{{\rm d}n_\gamma}{{\rm d}\nu}\, (1+z)^3\, c\, \sigma_\nu\, {\rm d} \nu,
 \end{equation}
where ${\rm d}n_\gamma/{\rm d}\nu$ is the spectral distribution of the background ionizing photons, 
$c$ is the speed of light, and $\sigma_\nu = \sigma_0\, (\nu/\nu_0)^{-3}$ with 
$\sigma_0 = 6.3 \times 10^{-18} \cm^2$ and $\nu_0$ being the frequency of hydrogen ionization threshold.
Assuming a power law spectral distribution of the form 
${\rm d}n_\gamma/{\rm d}\nu = (n_\gamma^0/\nu_0) (\nu/\nu_0)^{-\eta-1}$, 
in which $n_\gamma^0$ is related to the total photon number density $n_\gamma$ by 
$n_\gamma = n_\gamma^0/\eta$, then the HI photoionization rate can be written as
\begin{equation}\label{Gamma_HI}
\Gamma_{\rm HI} \;=\; \frac{\eta}{\eta+3}\, n_\gamma\, (1+z)^3\, c\, \sigma_0.
\end{equation}
In the following we assume $\eta = 3/2$ to approximate the spectra of starburst galaxies 
\citep{2005MNRAS.363.1031F}.

\begin{figure}[t]
\centering{
\includegraphics[scale=0.4]{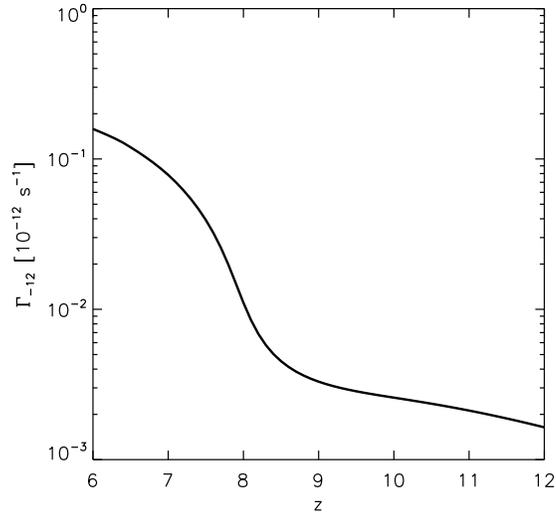}
\caption{The redshift evolution of the hydrogen ionization rate $\Gamma_{\rm -12}$.}
\label{Fig.Gamma_12}
}
\end{figure}

It has been suggested that the characteristic length $\lambda_0$ in Eq.(\ref{Eq.mfp}) 
is related to the Jeans length and can be fixed by comparing
with low redshift observations \citep{2005MNRAS.361..577C,2013ApJ...772...93K}.
We take $\lambda_0 = A_{\rm mfp} r_{\rm J}$, where $r_{\rm J}$ is the physical Jeans length.
Taking the proportional constant $A_{\rm mfp}$ as a free parameter, the comoving number density
of background ionizing photons $n_\gamma$, or equivalently the HI photoionization rate 
$\Gamma_{\rm HI}$, can be solved by combining Eq.(\ref{n_gamma}) - (\ref{critical_density}) 
and Eq.(\ref{Gamma_HI}). We
scale the hydrogen photoionization rate to be $\Gamma_{\rm HI} = 10^{-12.8} \psec$ at redshift 6, as 
suggested by recent measurements from the Ly-$\alpha$ forest 
\citep{2011MNRAS.412.1926W,2011MNRAS.412.2543C}. Then the parameter $A_{\rm mfp}$ is 
constrained to be $A_{\rm mfp} = 0.482$.
The redshift evolution of the hydrogen photoionization rate due to the ionizing background
is shown in Fig.~\ref{Fig.Gamma_12}.
Note that by scaling the background photoionization rate of hydrogen 
to the observed value, we implicitly take into account the possible absorptions 
due to minihalos and low column density systems.

In the above treatment of the ionizing background, the derived intensity is effectively 
the averaged value over the whole Universe. Due to the clustering of the ionizing sources, however,
the ionizing background should fluctuate significantly from place to place at the end of reionization.
The detailed space fluctuations of the ionizing background would be challenging to incorporate, and
for the purpose of illustrating the island model and predicting the statistical results in the next section,
here we use a uniform ionizing background with the averaged intensity.

\section{The Island model of Reionization}\label{results}
\subsection{Ionization at the surfaces of neutral islands}

We now use the excursion set model developed above to study the neutral islands
during the reionization process. In the section \ref{toy_model}, we used a simple toy model 
to illustrate the basic formalism, but we have noted that it is based on 
an unrealistic assumption,  
that the ionizing photons permeate through the neutral islands. 
Here we consider more physically motivated model assumptions.

We assume that a spatially homogeneous 
ionizing background flux is established throughout all of the ionized 
regions at redshift $z_{\rm back}$.  
These ionizing photons can not penetrate the neutral islands, but were consumed
near the surface of the islands. 
We may then assume that the photons consumed by an island at any instant
is proportional to its surface area, or in terms of mass, $M^{2/3}$.
The number of background ionizing photons consumed is then given by
\begin{equation}\label{Nph}
N_{\rm back} = \int F(z)\,\Sigma_{\rm I} (t) \,{\rm d}t,
\end{equation}
where $\Sigma_{\rm I}$ is the physical surface area of the neutral island, 
while $F(z)$ is the physical number flux of background ionizing photons which is related to the
comoving photon number density by $F(z)=n_\gamma(z)\, (1+z)^3\,c/4$. 
For spherical islands, the surface area is
related to the scale radius by $\Sigma_{\rm I}=4\pi R^2 / (1+z)^2$, in which $R$ is in comoving coordinates. 
For non-spherical islands, 
one could still introduce a characteristic scale $R$ and 
the area would be related to $R^2$. In fact,
under the action of the ionizing background, non-spherical 
neutral regions have a tendency to evolve to
spherical ones because a sphere has the minimum surface area for the same volume.

\begin{figure}[t]
\centering{
\subfigure{\includegraphics[scale=0.4]{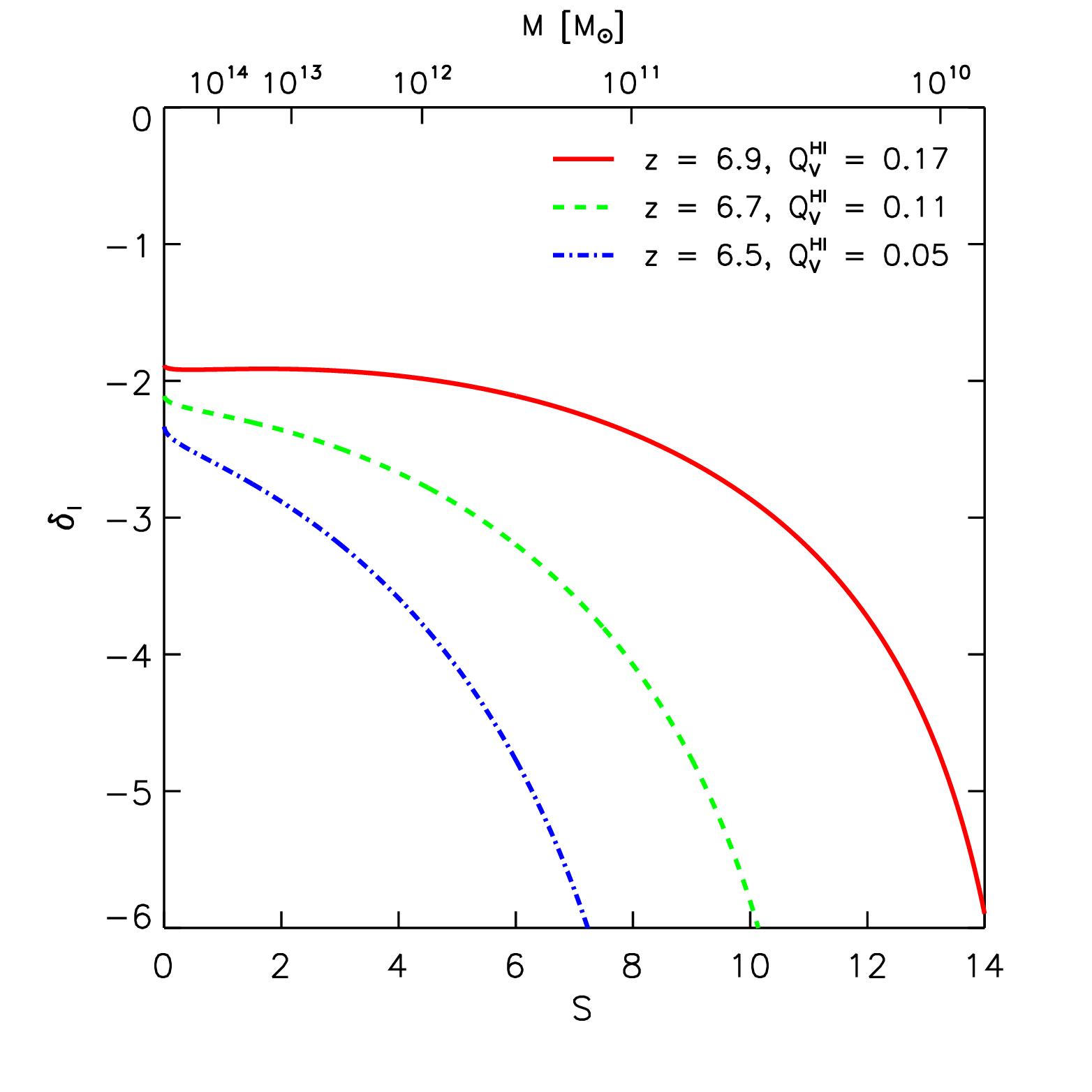}}
\subfigure{\includegraphics[scale=0.4]{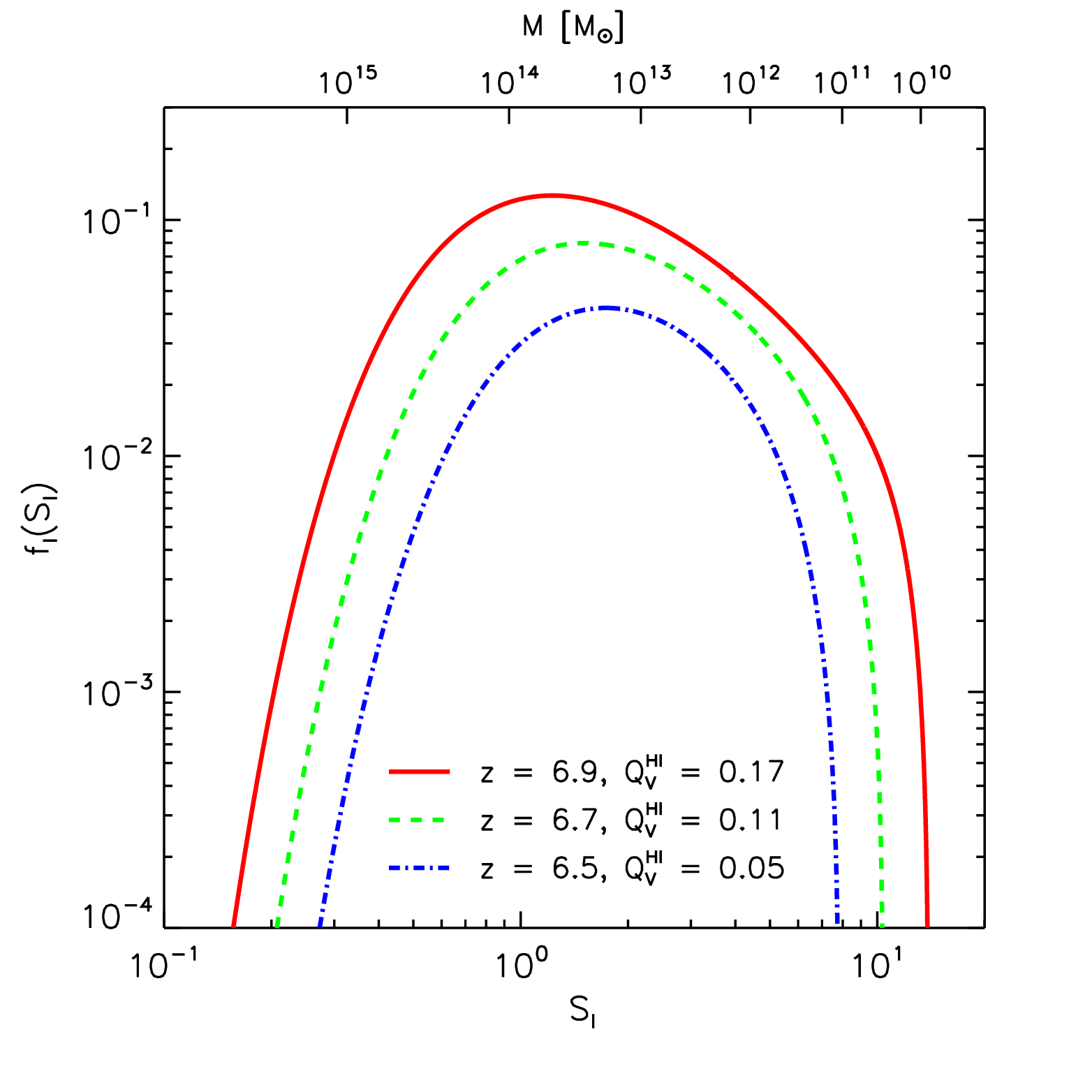}}
\caption{{\it Left panel}: the island barriers for our fiducial model. 
The solid, dashed and dot-dashed curves are for redshifts 6.9, 6.7 and 6.5 
from top to bottom respectively, and the corresponding neutral fractions of the Universe
(excluding the bubbles in islands)
are $Q_{\rm V}^{\rm HI} = $ 0.17, 0.11, and 0.05, respectively.
{\it Right panel}: The corresponding first down-crossing distributions at the
 same redshifts as the left panel.}
\label{Fig.barrier_fI_vS}
}
\end{figure}

The usual excursion set approach does not contain time or history, 
and everything is determined from the information at a given redshift. 
However, we see from Eq.~(\ref{Nph}) that the consumption of the 
ionizing background photons by an island depends on its history.
Below we try to solve this problem by considering some simplified assumptions.
We assume that the neutral islands shrink with time, 
and the hydrogen number density around an island is nearly a constant, which is approximately 
true when we are considering large scales. 
For simplicity, let us consider a spherical island. 
When the island shrinks, counting the required number of ionizations gives
\begin{equation}
n_{\rm H}(R)(1+\bar{n}_{\rm rec}) \,4\pi R^2\, (-{\rm d}R)\,  = 
F(z)\,\frac{4\pi R^2}{(1+z)^2}\, {\rm d}t,
\end{equation}
where the hydrogen number density $n_{\rm H}$ is in comoving coordinates, so that
\begin{equation}
\frac{{\rm d}R}{{\rm d}t} = -\frac{F(z)/(1+z)^2}{n_{\rm H}(R) (1+\bar{n}_{\rm rec})}
\approx -\frac{F(z)/(1+z)^2}{\bar{n}_{\rm H} (1+\bar{n}_{\rm rec})}.
\end{equation}
Integrating from the background onset redshift $z_{\rm back}$ to redshift $z$, we have  
\begin{equation}\label{eq.dR}
\Delta R \equiv R_i - R_f = \int_z^{z_{\rm back}} \frac{F(z)}
{\bar{n}_{\rm H} (1+\bar{n}_{\rm rec})} \, \frac{{\rm d}z}{H(z)(1+z)^3},
\end{equation}
where $R_i$ and $R_f$ denote the initial and final scale of the island respectively.
This shows that the change in $R$ is independent of the mass of the island, but 
depends solely on the elapsed time.
The total number of background ionizing photons consumed is given by
\begin{equation}\label{Eq.DeltaN}
N_{\rm back} = \frac{4\pi}{3}  \left(R_i^3-R_f^3\right) \bar{n}_{\rm H} (1+\bar{n}_{\rm rec}),
\end{equation}

\subsection{Island Size Distribution}

With this model for the consumption behavior of the background ionizing photons, and
taking the fiducial set of parameters, we plot the island barriers of 
inequation (\ref{Eq.islandBarrier})
in the left panel of Fig.~\ref{Fig.barrier_fI_vS} for several redshifts.
The corresponding first down-crossing distributions as a function of the host island scale 
$S_{\rm I}$ (i.e. including ionizing bubbles inside
the island) are plotted in the right panel of Fig.~\ref{Fig.barrier_fI_vS}. 

Unlike the toy model with permeating ionizing photons, in this model the shape of the 
island barriers is drastically different from the bubble barriers, hence a 
different shape of the first down-crossing distribution curves.
The island and bubble barriers have the same intercept at $S \sim 0$, because
on very large scales, the contribution of the ionizing background which is proportional
to the surface area would become unimportant when compared with the self-ionization which is
proportional to the volume. However, the island barriers bend downward at $S>0$, because
of the contribution of the ionizing background. As the barrier curves become gradually steeper
when approaching larger $S$, 
it is increasingly harder for the random walks to first down-cross them at smaller scales, 
even though on the smaller scales the dispersion of the random trajectory grow larger. 
As a result, the first down-crossing distribution rapidly increases to a peak value and drops 
down on small scales, and there 
is a mass-cut on the host island scale, $M_{\rm I,min}$, at each redshift in 
order to make sure $K(M,z)\ge 0$. This lower cut on the island mass scale 
assures $\Delta R \le R_i$, i.e. the whole island is not completely ionized during 
this time by the ionizing background, and $M_{\rm I,min}$ is the minimum mass of the host 
island at $z_{\rm back}$ that can survive till the redshift $z$ under consideration.

\begin{figure}[t]
\centering{
\includegraphics[scale=0.4]{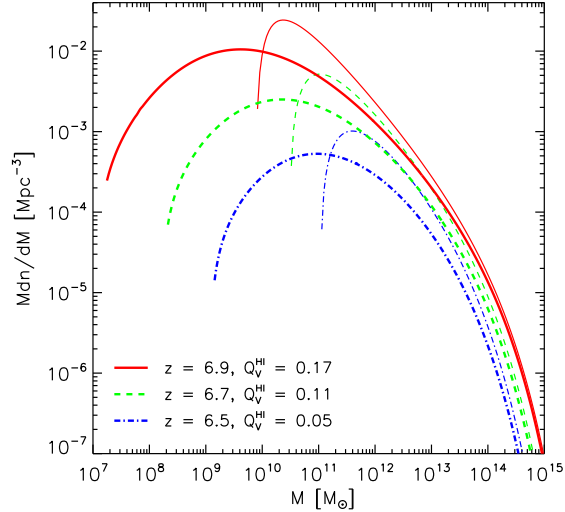}
\caption{The mass function of the host islands in terms of the mass at redshift $z$ (thick lines)
and the initial mass at redshift $z_{\rm back}$ (thin lines) for our fiducial model.
The solid, dashed, and dot-dashed lines are for $z=$ 6.9, 6.7, 
and 6.5, from top to bottom respectively.}
\label{Fig.MF_host_vS}
}
\end{figure}

The mass distribution function of the host islands can be obtained directly from 
Eq.~(\ref{eq.hostMF}), from which we 
can see clearly the shrinking process of these islands.
What we are interested is the mass of the host island at 
redshift $z$, but the mass scale $M$ in 
Eqs.~(\ref{Eq.IslandCondition}-\ref{Eq.K}) is the initial island 
mass at redshift $z_{\rm back}$. We may convert the two masses
using Eq.~(\ref{eq.dR}):
\begin{equation}
\frac{M_f}{M_i}=(1-\frac{\Delta R}{R_i})^3 
\label{eq.Mf}
\end{equation}
Islands with initial radius $R_i<\Delta R $ would not survive, and islands with larger 
radius would also evolve into smaller ones.

The distributions of the host island mass (including ionized bubbles inside) 
are plotted for $z = $ 6.9, 6.7, and 6.5 in Fig.~\ref{Fig.MF_host_vS} as thick 
lines. The distributions of the corresponding progenitors at redshift $z_{\rm back}$ are plotted as thin lines.
Using our fiducial model parameters, the volume filling factors of these progenitors 
at $z_{\rm back}$ are $Q^{\rm host}_{\rm V,i} = $ 0.51, 0.31, and 0.14, for the host islands 
survived at $z = $ 6.9, 6.7, and 6.5 respectively.
The initial mass distribution of these progenitors all have
a very steep lower mass cutoff, because below that minimal mass by redshift $z$ 
the whole island would be completely ionized by the background photons. Due to the mapping 
of Eq.~(\ref{eq.Mf}), the cutoff in the final mass distribution is not as sharp as the 
initial mass distribution and the whole distribution curve begin to bend down at lower masses.

\subsection{Bubbles-in-Islands}

However, the total mass function of the host islands does not give a full 
picture of the reionization process, since there could be ionized bubbles inside these 
islands. Even though the outside ionization background is shielded from 
the center of the neutral islands, there might be
galaxies formed inside the neutral islands, and the photons 
emitted by these galaxies ionize part of
the islands. The neutral islands are located in underdense regions, so fewer 
galaxies formed, nevertheless, by the end of the epoch of reionization, 
galaxy formation inside them can not be neglected. 

\begin{figure}[t]
\centering{
\subfigure{\includegraphics[scale=0.4]{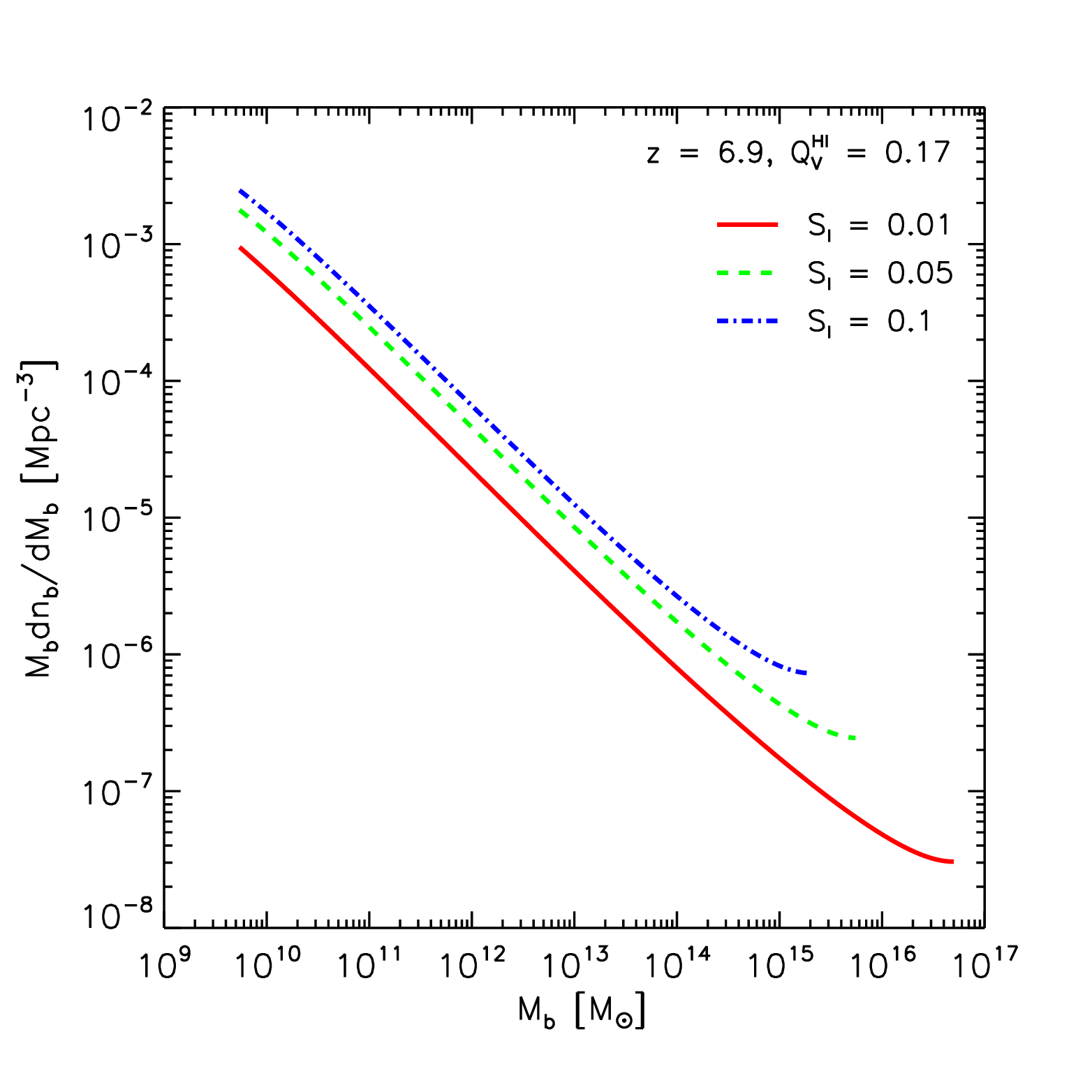}}
\subfigure{\includegraphics[scale=0.4]{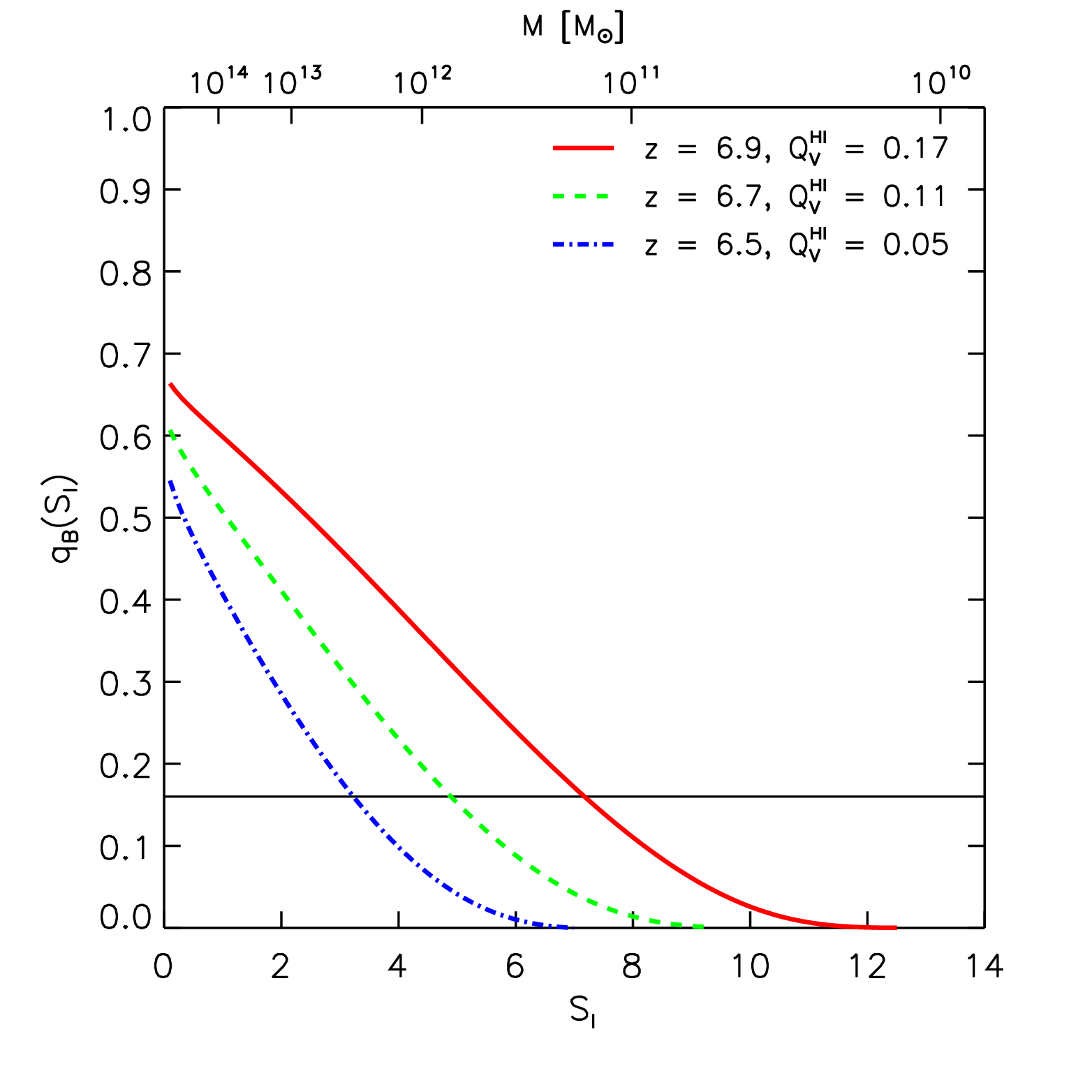}}
\caption{{\it Left panel}: The mass function of bubbles in an island 
of scale $S_{\rm I} = $ 0.01, 0.05, and 0.1, from bottom to top respectively. 
The redshift shown here is 6.9. {\it Right panel}: The average mass fraction 
of bubbles in an island as a function of the island scale at redshifts $z = 6.9$, 
6.7, and 6.5, from top to bottom respectively. The percolation threshold $p_c = 0.16$ 
is also shown as the horizontal line.}
\label{Fig.qB_vS}
}
\end{figure}

As discussed in \S \ref{bubbles_in_island}, the distribution of bubbles in an island can be 
calculated from the conditional probability of up-crossing the bubble barrier after down-crossing
the island barrier. We plot the resulting mass function of inside bubbles for three 
different host islands at redshift $z=6.9$ in the left panel of Fig.~\ref{Fig.qB_vS}. 
The masses of the host islands are $M\approx 2\times 10^{17} M_\odot \;(S_{\rm I}=0.01)$, 
$2\times 10^{16} M_\odot \;(S_{\rm I}=0.05)$, and $8\times 10^{15} M_\odot \;(S_{\rm I}=0.1)$, 
from bottom to top respectively. We see the bubbles in islands follow a power law 
distribution, with small bubbles more numerous. The upward trend at the large scale end
on each mass distribution curve is due to the numerical error in the up-crossing probability when the
inside bubble scale approaches the host island scale.

To assess the total amount of bubbles in islands, we plot in the right panel of 
Fig.~\ref{Fig.qB_vS} the average mass fraction 
of bubbles-in-island as a function of the host island mass. 
We see that there could be a sizable fraction of the host island which is ionized from within, 
especially for the larger islands. 
At $z=6.9$ and for $M>10^{12} M_\odot$, this fraction 
is higher than 35\%, and it is higher than 60\% for $M>10^{14} M_\odot$ host islands 
at the same redshift, so within these large neutral islands smaller ionized bubbles flourishes. 
From the excursion set point of view, it is not unusual for the random trajectory 
to turn upward the bubble barrier after just down crossed the island barrier, 
especially at large scales where the displacement between the island barrier 
and the bubble barrier is small. Therefore, even though the whole region is underdense, a large
fraction of it could be sufficiently dense for galaxies to form and create
ionized regions around them. The bubble fraction drops sharply for smaller 
islands, because the island barrier departs from the bubble barrier rapidly at small scales, and
it is less likely to form galaxies inside small islands with very low densities. 
Interestingly, as redshift decreases this fraction drops down. 
For $z=6.5$, it is about 7\% 
for $M\sim10^{12} M_\odot$ host islands, 
and about 42\% for $M\sim10^{14} M_\odot$ host islands.
This is because what are left at later time are relatively deep underdense regions, 
and the probability of forming galaxies in such underdense environments is lower.

\begin{figure}[t]
\centering{
\subfigure{\includegraphics[scale=0.4]{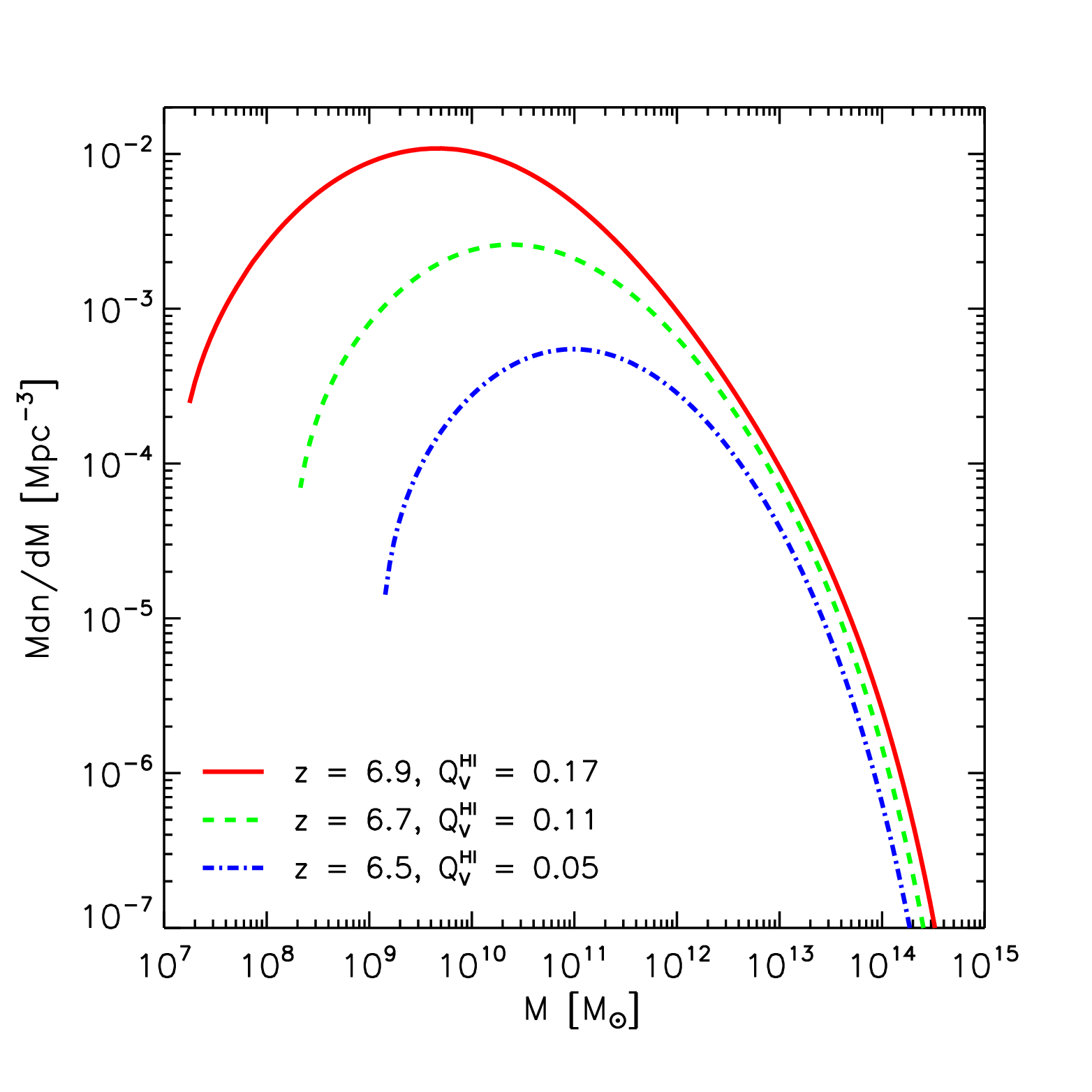}}
\subfigure{\includegraphics[scale=0.4]{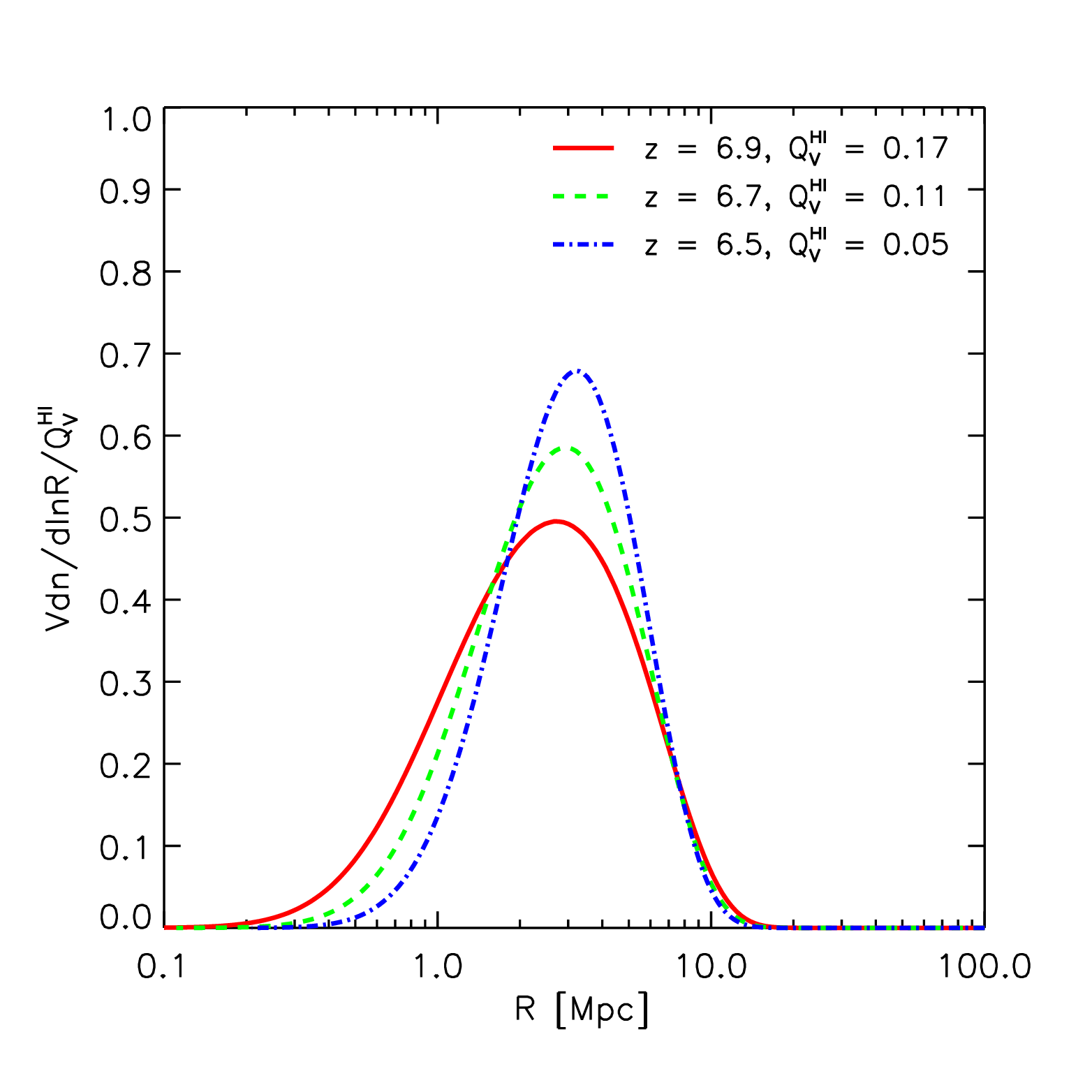}}
\caption{{\it Left panel}: The mass function of neutral islands at redshift $z = $ 
6.9, 6.7, and 6.5, from top to bottom respectively. The corresponding volume 
filling factor of the neutral islands at these redshifts are $Q_{\rm V}^{\rm HI} = $ 0.17, 
0.11, and 0.05, respectively. {\it Right panel}: The size distribution of neutral 
islands, with the scale $R$ converted from their volume, at redshifts $z = 6.9$, 
6.7, and 6.5, from bottom to top at the center respectively.}
\label{Fig.MF_Rdis_vS_nopc}
}
\end{figure}

Excluding the bubbles in islands, we plot the mass function and the size 
distribution of the {\it net}
neutral islands in the left and right panel of Fig.~\ref{Fig.MF_Rdis_vS_nopc} respectively. 
The solid, dashed, and dot-dashed lines are for $z = $ 6.9, 6.7, and 6.5, with a 
volume filling factor of the net neutral islands of 
 $Q_{\rm V}^{\rm HI} = 0.17 (z = 6.9), 0.11 (z = 6.7)$, and 0.05 ($z = 6.5$), respectively. 
Similar to the host island mass function shown in Fig.~\ref{Fig.MF_host_vS}, 
there is also a small scale cutoff on the neutral island mass due to the existence of
an ionizing background. Because of the high bubbles-in-island fraction 
in large host islands, excluding the 
bubbles in islands results in much fewer large islands. As seen from the size 
distribution in the right panel, in which the scale $R$ is converted from the 
neutral island volume assuming spherical shape, both the mass fractions of large 
and small islands decrease with time, and the distribution curve becomes sharper 
and sharper, but the characteristic scale of the neutral islands remains almost unchanged.

Fig.~\ref{Fig.MF_Rdis_vS_nopc} shows basically the number and mass distribution 
of the neutral components of 
the host islands. However, the results of bubbles-in-island fraction in the 
right panel of Fig.~\ref{Fig.qB_vS}
 show that within large host islands, a large fraction  
of the island volume could be ionized by the photons from newly formed galaxies within. 
A naive application of the host island mass function may greatly overestimate the mean
neutral fraction of the Universe, while the application of the neutral island size distribution, 
as shown in the right panel of Fig.~\ref{Fig.MF_Rdis_vS_nopc}, 
would never reveal the real image of the ionization field. 
Indeed, if there are so many ionized bubbles inside large neutral islands,
it may be difficult to visually identify the host islands. In light of this, we need 
to consider the condition under which the isolated island picture is still applicable. 
Especially, if the bubbles inside an island are so numerous and large as to overlap with 
each other, they may form a network which percolates through the whole island, 
and break the island into pieces, or form a sponge-like topology of neutral and ionized regions.

\subsection{Percolation Model}

Within the spherical model, it is difficult to deal with the sponge-like 
topology, but we may limit ourselves to the case where the treatment is still valid.
According to the theory of percolation, in a binary phase system, percolation of one phase 
occurs when the filling factor of it exceeds
a threshold fraction $p_c$ (see e.g. \citealt{1991fds.book.....B}). 
In the context of cosmology, \citet{1993ApJ...413...48K} obtained the percolation 
threshold $p_c$ for the clustered large scale structures from cosmological simulations. 
However, the spatial distribution of ionized bubbles and neutral islands are much less 
filamentary than the gravitationally clustered dark matter or galaxies. As the 
ionization field follows the density field \citep{2012arXiv1211.2832B}, which is 
almost Gaussian on large scales \citep{2013arXiv1303.5084P}, here we use the 
percolation threshold for a gaussian random field of $p_c = 0.16$ 
\citep{1993ApJ...413...48K}, below
 which we may assume that the bubbles-in-island does not 
percolate through the whole island.

The problem of percolation appears in several stages of reionization. 
At the early stage of reionization, the filling factor of ionized bubbles increases 
as the bubble model predicted. Once the bubble filling factor becomes larger than 
the percolation threshold $p_c$, the ionized bubbles are no longer isolated, 
and the predictions made from the bubble model are not accurate anymore. 
Therefore, the threshold $p_c$ sets a critical redshift $z_{\rm Bp}$, below which 
the bubble model may not be reliable. Similarly, the model of neutral islands can make
accurate predictions only below a certain redshift $z_{\rm Ip}$, when the
island filling factor is below $p_c$. The ionizing background was set up after 
the ionized bubbles percolated but before the islands were all 
isolated, so $z_{\rm Bp} > z_{\rm back}>z_{\rm Ip}$.
Finally, the percolation threshold may also be applied to the bubbles-in-island fraction. 
An island with a high value of $q_{\rm B}$ may not qualify 
as a whole neutral island, and the bubbles inside it are probably not isolated regions.

\begin{figure}[t]
\centering{
\includegraphics[scale=0.4]{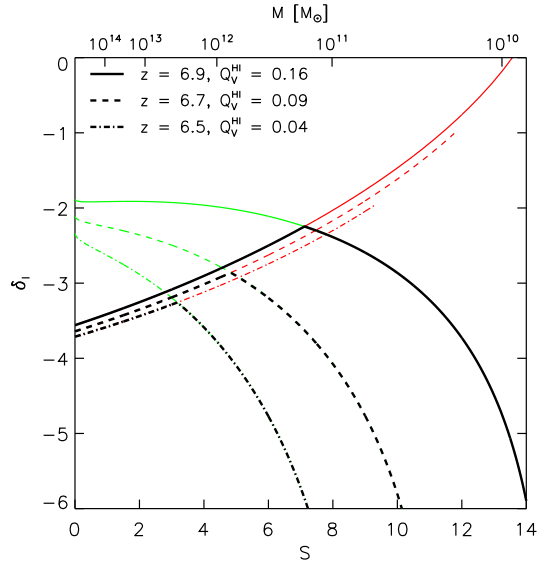}
\caption{The basic island barriers (green curves), the percolation threshold 
induced barriers (red curves), and the effective island barriers (black curves)
for our fiducial model. The solid, dashed and dot-dashed curves 
are for redshifts 6.9, 6.7 and 6.5 from top to bottom respectively.}
\label{Fig.barriervS}
}
\end{figure}

It may be desirable to consider also the distribution of 
those {\it bona fide} neutral islands, for which the bubble fraction is below 
the percolation threshold, i.e. after excluding those islands 
with $q_{\rm B} > p_c$. This percolation criterion of $q_{\rm B} < p_c$ acts 
as an additional barrier for finding islands, 
those islands with high bubbles-in-island fractions are excluded, 
but the neutral regions in them contribute to the number of smaller islands.
This additional barrier is obtained by solving $q_{\rm B}(S_{\rm I},\delta_{\rm I};z) < p_c$,  
and are plotted in Fig.~\ref{Fig.barriervS} with red lines for redshift $z = $
6.9, 6.7, and 6.5, from top to bottom respectively. The basic island barriers are also plotted
in the same figure with green lines. The combined effective island barriers 
are shown as black lines. 
The barrier resulted from the percolation criterion takes its effect on large scales 
as larger islands could have higher bubbles-in-island fractions, and 
larger scale islands need to be more underdense to keep the whole region mostly neutral.
The basic island 
barrier (\ref{Eq.islandBarrier}) is effective on small scales, because small 
islands are easier to be swallowed by the ionizing background.
According to the percolation criterion, the island model can be reasonably 
applied at redshifts below $z_{\rm Ip} \sim 6.9$ in our fiducial model, though for 
other parameter set the value would be different.

\begin{figure}[t]
\centering{\includegraphics[scale=0.4]{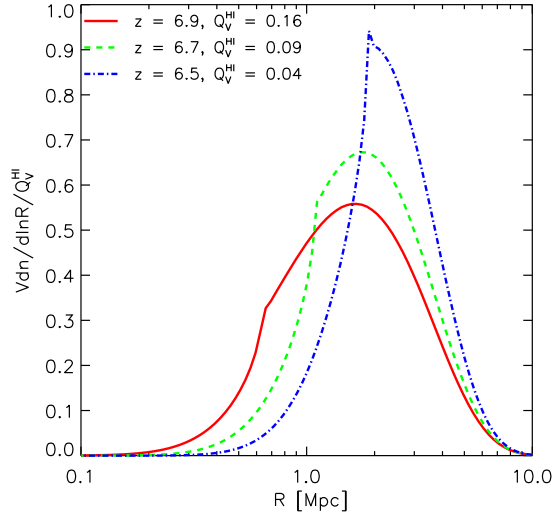}
\caption{The size distribution of neutral islands in our fiducial model taking into account the bubbles-in-island effect and the $p_c$ cutoff on bubbles-in-island fraction. The solid, dashed, and dot-dashed curves are for redshifts $z = 6.9$, 6.7, and 6.5, respectively, and the corresponding volume filling factors of neutral islands are 
$Q_{\rm V}^{\rm HI} =$ $0.16\; (z = 6.9)$, $0.09\; (z = 6.7)$, and $0.04\;  (z = 6.5)$, respectively. }
\label{Fig.Rdistr_vS}
}
\end{figure}

With the combined island barrier taking into account the bubbles-in-island effect, 
we find host islands by computing the first down-crossing distribution, and find 
bubbles in them by computing the conditional first up-crossing distribution with 
respect to the bubble barrier. Subtracting the bubbles in islands, the
mass distribution of the neutral islands and the volume filling factor of the neutral 
components  $Q_{\rm V}^{\rm HI}$ are obtained.
The resulting size distribution of the neutral islands in terms of the effective 
radii is plotted in Fig.~\ref{Fig.Rdistr_vS} for redshifts $z = 6.9$, 6.7, and 6.5. 
The distribution curve is normalized by the total neutral fraction in each redshift, 
which is 
$Q_{\rm V}^{\rm HI} = 0.16 (z = 6.9), 0.09 (z = 6.7)$, 
and 0.04  $(z = 6.5)$, respectively.

We note that after applying the $p_c$ cutoff, the resulting neutral fraction at 
a specific redshift differs a little from the model without the $p_c$ cutoff. 
Intuitively, the percolation threshold act only as a
different definition of islands, and should not change the ionization state of the IGM. This is true because 
those islands excluded by the percolation threshold will be considered as pieces of smaller islands that 
still contribute to the total neutral fraction. However, two competitive facts are taking effects in our 
island-finding procedure, which could make the results different. 
First, we have assumed that the 
bubbles in islands are all ionized, but neglected those small islands that 
could possibly exist in these relatively large bubbles.
When applying the $p_c$ cutoff, some large islands with large bubbles are excluded, and the random walk
would continue to enter the scales smaller than the bubbles, and could possibly find smaller islands that 
are embedded in large bubbles. Therefore, the model with $p_c$ cutoff could find more small islands that
are not accounted for in the model without $p_c$ cutoff, and tends to predict higher neutral fraction.
On the other hand, one large island with high bubbles-in-island fraction is taken as several smaller islands 
in the model with $p_c$ cutoff, and small islands are more significantly influenced by the ionizing background.
This fact would result in lower neutral fraction for the model with $p_c$ cutoff.
As the redshift decreases, more and more small islands are swallowed by the ionizing background, so 
the second effect gradually dominates over the first one.
With the fiducial parameters used here, the second effect dominates for the redshifts of interest, 
and the neutral fractions predicted in the model with $p_c$ cutoff is slightly lower than in the model without
$p_c$ cutoff.

As shown in Fig.~\ref{Fig.Rdistr_vS}, in this model, the island size distribution after $z_{\rm Ip}$ also has a peak.
For this set of model parameters, the characteristic size of neutral 
islands at $z=6.9$ is about 1.6 Mpc, 
but the distribution extends a range, with the lower value as small as 0.2 Mpc, 
and the high value as large as 10 Mpc. 
As the redshift decreases, small islands disappear rapidly because of the ionizing background.
This is qualitatively consistent with simulation results \citep{2008ApJ...681..756S} 
in which small islands are much rarer during the late reionization as compared to 
those small ionized bubbles in the early stage.
As the reionization proceeds, the large islands shrink and the small islands are being swallowed 
by the ionizing background, with the small ones disappearing more rapidly, and 
the peak position of the distribution curve shifts slightly towards larger scale but does not change much.
Due to the rapidly decreasing number of 
small islands, the distribution curve becomes narrower. 
The distribution also becomes taller with decreasing redshift because it is normalized against the 
volume neutral fraction $Q_{\rm V}^{\rm HI}$ at each redshift. 
With $Q_{\rm V}^{\rm HI}$ decreasing, the normalized distribution has narrower 
and higher peaks, but the absolute number of
neutral islands per comoving volume is decreasing.

\section{Conclusion}\label{Discuss}

This paper is devoted to the understanding of the late stage of the epoch of
reionization. According to the bubble model \citep{2004ApJ...613....1F} 
and radiative transfer simulations, reionization started with the ionization of
regions with higher-than-average densities, as stars and galaxies formed earlier in 
such regions, while the regions with lower average densities remained neutral 
for longer time. Inspired by the bubble model, here we try to understand 
the evolution of the remaining large neutral regions
which we call ``islands'' during the late stage of reionization. 
We developed a model of their mass distribution and evolution 
based on the excursion set theory. The excursion set theory is appropriate 
for constructing the ionized bubble model
and the neutral island model because the reionization field 
follows the density field on large scales \citep{2012arXiv1211.2821B}.

With the inclusion of an ionizing background, 
which should exist after the percolation of ionized regions, 
we set an island barrier on the density contrast in the excursion set theory for 
the islands to remain neutral, and an island was identified 
when the random walk first-{\it down}-crosses 
the island barrier.  We presented algorithms 
for computing the first-down-crossing distribution, obtained mass function for the islands,
and also provide a semi-empirical way to determine the intensity 
of the ionizing background during the late reionization era.

We first illustrated the formalism of computation with 
a simple toy model, where the number of 
consumed ionizing background photons per unit time is proportional to the volume of the 
island, i.e. the ionizing background is uniformly distributed within the island. While this 
is not realistic, it is relatively simple to derive the analytical expression of the 
neutral island mass function. The model predicts a large number 
of small islands. We then considered a more realistic model, where the ionizing background 
only causes the ionization at the surface of the island, so that the consumption rate of 
the ionizing background is proportional to the surface area of the island. 
Under the action of such ionizing photons, an island would shrink with time. 
The larger islands shrink, while smaller ones disappear. As a result of this, 
there is a minimal initial mass at the ``background onset redshift'' for the islands.
We obtained the distribution function of the initial and final mass of the islands at
different redshifts. 

However, ionized bubbles also formed within the large neutral islands, these 
bubbles-in-islands must be take into account. For this we considered two barriers, 
the island barrier and the bubble barrier, at the same time. The former inludes the effect
of ionizing background at the surface of the island, while the latter does not.
The bubbles embedded in an island were found by computing the first-{\it up}-crossings 
over the bubble barrier after the random walks have {\it down}-crossed 
the island barrier at the host island scale, and the volume fraction of 
bubbles-in-island are obtained. We find that for a large island, a large portion of
its interior could be ionized. 

The bubbles-in-island problem limited the applicability of this model, because in
non-symmetrical cases, the presence of bubbles may break the island into small
pieces, which would increase the exposed surface of the island. 
To address this problem, we applied a percolation 
criterion as an additional island barrier on large scales. Islands with large
bubbles-in-island fraction are excluded, because in the real world where the
bubbles are not spherical and concentric, these bubbles would have percolated through
the island and break it into smaller islands.
Using the combined island barrier and excluding the ionized bubbles in the islands, 
the volume filling factor of neutral islands in the Universe
and the size distribution of the neutral islands were derived. Our island model applies 
to the large scale structure of neutral regions in the linear regime, but 
it may be possible to account for the small scale physics, 
such as the minihalo absorptions, by introducing a consuming term in 
the formula (e.g. \citealt{2005MNRAS.363.1031F,2012ApJ...747..127Y}).

At a given instant shortly after the isolation of islands, our model predicts
that the size distribution of the islands has a peak of a few Mpc, 
depending on the model parameters.  
As the redshift decreases, the small islands disappear rapidly 
while the large ones shrinks, but the characteristic 
scale of the islands does not change much.
Eventually, all these large scale
neutral islands are swamped by ionization, only compact neutral regions such as galaxies
or minihalos remain.

In our semi-empirical model of the ionizing background, the main absorbers of the ionizing photons 
are self-shielded Lyman limit systems. However, one needs to check to what extent the lower density 
neutral islands regulate the mean free path of the ionizing photons. The mean free path due to the 
existence of islands can be estimated with $\lambda_{\rm mfp}^{\rm I}(z) \sim 1/[\int \pi R_{\rm f}^2\,
({\rm d}n_{\rm f}/{\rm d}M_{\rm f})\, {\rm d}M_{\rm f}]$, where $R_{\rm f}$ and 
${\rm d}n_{\rm f}/{\rm d}M_{\rm f}$ are the size and mass function of final host islands respectively at 
redshift $z$. We found that at $z=6.9$, the mean free path of ionizing photons due to islands 
$\lambda_{\rm mfp}^{\rm I} \sim 1.12$ physical Mpc as compared with that due to Lyman limit system 
$\lambda_{\rm mfp} \sim 0.30$ physical Mpc. At $z=6.7$, $\lambda_{\rm mfp}^{\rm I} \sim 2.68$ 
physical Mpc as compared with $\lambda_{\rm mfp} \sim 0.38$ physical Mpc, while $z=6.5$, 
$\lambda_{\rm mfp}^{\rm I} \sim 7.93$ physical Mpc as compared with $\lambda_{\rm mfp} \sim 0.48$
 physical Mpc. Therefore, the mean free path of ionizing photons due to islands is always much larger than
the mean free path due to Lyman limit systems, and the effect of islands on the ionizing background is 
negligible as compared to the effect of small scale dense clumps. As the redshift decreases, the large
 scale islands become less and less important in regulating the mean free path of ionizing photons.
Considering the dominant contribution of the Lyman limit systems to the IGM opacity, would they also 
contribute significantly to the neutral volume during the late era of reionization?
The volume fraction of these Lyman limit systems can be estimated with 
$1\,-\, F_{\rm V}(\Delta_{\rm crit})$, and it is about 0.0062, 0.0046, and 0.0036, respectively for $z = 6.9$,
$6.7$, and $6.5$, much lower than the volume filling factor of islands. 
Because of the much lower number density and larger size of islands, the mean free path due to islands is 
much larger than that due to Lyman limit systems, even though the volume filling fraction of islands is larger. 
Therefore, the majority of neutral volume of the IGM is occupied by the islands, which is consistent with 
our model assumption, but the opacity of the IGM is dominated by the dense Lyman limit systems.

The results shown here are primarily qualitative, the quantitative 
predictions are dependent on our model assumptions and model parameters. Current 
observations have not yet been able to constrain such
parameters effectively, and they can be redshift-dependent. 
Our model assumption may also be too simplistic, 
for example, we may over-predict the number of large islands because 
they are more likely non-spherical, 
and the ionizing background should have stronger effect on them as they 
have larger surface area for the same volume.
These uncertainties could be constrained in the future if the model
predictions are compared with 21cm and/or other observations, and  
as the properties of ionizing sources, the evolution 
of neutral islands, and the intensity of the ionizing background become
better known. We shall investigate the late reionization epoch by 
numerical simulations and compare it with the analytical models
in subsequent works.

\acknowledgments
We deeply appreciate the insight of the referee and the constructive comments.
We thank Jun Zhang, Jie Zhou, Hy Trac and Renyue Cen 
for many helpful discussions. This work is supported by 
the Ministry of Science and Technology 863 project grant 2012AA121701, the 
NSFC grant 11073024, and the John Templeton foundation. 
Y.X. is supported by China Postdoctoral Science Foundation and 
by the Young Researcher Grant of National Astronomical Observatories, 
Chinese Academy of Sciences.
Support for the work of M.S. was provided by NASA through Einstein Postdoctoral 
Fellowship grant number PF2-130102 awarded by the Chandra X-ray Center, which is 
operated by the Smithsonian Astrophysical Observatory for NASA under contract NAS8-03060.
Zuihi Fan is supported by NSFC under grant 11173001 and 11033005.

\bibliography{references}
\bibliographystyle{hapj}

\end{document}